\documentclass[fleqn,usenatbib]{mnras}

\usepackage{amssymb}
\usepackage{mathrsfs}
\usepackage{xspace}

\usepackage[T1]{fontenc}
\usepackage{ae,aecompl}

\newcommand{\rsun}{\,\mbox{$\rm R_{\odot}$}\xspace}

\newcommand{\ionjb}[2]{#1{~\sc{\romannumeral #2}}}

\newcommand{\kms}{\hbox{km~s$^{-1}$}\xspace}
\newcommand{\ms}{\hbox{ms$^{-1}$}\xspace}

\newcommand{\vsini}{\hbox{$v$\,sin\,$i$}\xspace}

\newcommand{\degs}{$\degr$\xspace}

\newcommand{\ha}{H$\alpha$\xspace}

\newcommand{\hd}{HD~189733\xspace}
\newcommand{\hdone}{HD~189733b\xspace}
\newcommand{\hdtwo}{HD~209458b\xspace}

\newcommand{\cahk}{Ca~{\sc ii}~H~\&~K\xspace}
\newcommand{\nad}{Na~{\sc i}~D1~\&~D2\xspace}

\newcommand{\harps}{{\sc harps}\xspace}

\newcommand{\transiti}{Transit~1\xspace}
\newcommand{\transitii}{Transit~2\xspace}
\newcommand{\transitiii}{Transit~3\xspace}

\usepackage{graphicx}	
\usepackage{amsmath}	
\usepackage{amssymb}	

\title[Excess absorption during the transit of HD189733b]{The origin of the excess transit absorption in the \hbox{HD 189733} system: planet or star?}

\author[J.R.~Barnes et al.]
{J.R.~Barnes$^{1}$, 
C.A.~Haswell$^{1}$,
D.~Staab$^{1}$,
G.~Anglada-Escud\'{e}$^{2}$, \\
$^{1}$ Department of Physical Sciences, The Open University, Walton Hall, Milton Keynes MK7 6AA, UK. \\
$^{2}$ School of Physics and Astronomy, Queen Mary, University of London, 327 Mile End Rd. London, UK
}

\date{Received May 2016; Accepted for publication July 2016}

\pubyear{2016}

\begin{document}
\label{firstpage}
\pagerange{\pageref{firstpage}--\pageref{lastpage}}
\maketitle

\begin{abstract}
We have detected excess absorption in the emission cores of \cahk during transits of HD 189733b for the first time. Using observations of three transits we investigate the origin of the absorption, which is also seen in \ha and the \ionjb{Na}{1} D lines. Applying differential spectrophotometry methods to the \ionjb{Ca}{2} H and \ionjb{Ca}{2} K lines combined, using respective passband widths of \hbox{$\Delta\lambda =$} \hbox{$0.4\ \&\ 0.6$ \AA}\ yields excess absorption of ${ t_d =}\ 0.0074$\ $\pm$\ $0.0044$ ($1.7\sigma$; Transit 1) and \hbox{$0.0214$\ $\pm$\ $0.0022$} ($9.8\sigma$; Transit 2). Similarly, we detect excess \ha absorption in a passband of width $\Delta\lambda = 0.7$ \AA, with \hbox{${ t_d =\,0.0084\,\pm\,0.0016}$} ($ 5.2\sigma$) and \hbox{${ 0.0121\,\pm\,0.0012}$} ($ 9.9\sigma$). For both lines, Transit 2 is thus significantly deeper. Combining all three transits for the \ionjb{Na}{1} D lines yields excess absorption of \hbox{${ t_d =\,0.0041\,\pm\,0.0006}$} ($ 6.5\sigma$). By considering the time series observations of each line, we find that the excess apparent absorption is best recovered in the stellar reference frame. These findings lead us to postulate that the main contribution to the excess transit absorption in the differential light curves arises because the normalising continuum bands form in the photosphere, whereas the line cores contain a chromospheric component. We can not rule out that part of the excess absorption signature arises from the planetary atmosphere, but we present evidence which casts doubt on recent claims to have detected wind motions in the planet's atmosphere in these data.
\end{abstract}




\section{Introduction}
\label{section:intro}

Appropriately, the first known transiting exoplanet, \hbox{HD 209458}b \citep{charbonneau00transit,henry00hd209458}, also yielded the first detection of an atmosphere outside the solar system. The excess absorption of $(2.32 \pm 0.57) \times 10^{-4}$ in the 5893 \AA\ Na resonance doublet compared with neighbouring continuum passbands suggested that Na was present at a lower level than predicted \citep{seager00transmission}. Clouds, hazes and photo-ionisation of Na were investigated by \cite{seager03hd209458}, \cite{fortney03hd209458} and \cite{barman07} as possible causes for the low observed Na absorption. A complete optical transmission spectrum of \hdtwo\ spanning 3000\,-\,7000 \AA\ was observed for the first time with the Hubble Space Telescope (HST), additionally revealing strong Rayleigh scattering \citep{etangs08rayleigh} at wavelengths of 3000\,-\,5000 \AA. With a later spectral type of K1V\,-\,K2V, \hd causes lower incident radiation of the orbiting hot Jupiter despite its smaller orbital radius. HST transmission spectroscopy of \hdone\ have revealed that Na is observed as a weak feature \citep{pont08hd189733,pont13hd189733} while strong Rayleigh scattering dominates at shorter optical and UV wavelengths \citep{pont08hd189733,sing11hd189733,pont13hd189733}.

\begin{table*}
\begin{center}
\caption{HARPS observations during transit of \hbox{HD 189733b}. Columns 5-9 list S/N ratios for the \cahk, the mean of the nearby Mount Wilson S index V \& R continuum passbands, \ha and the mean of the \ha A \& B continuum passbands. \label{table1}}
\begin{tabular}{cccccccccc}
\hline
                 &             &                    &             &           &                        &                 & Passband S/N ratios   &              &                    \\
\cline{5-9}
\vspace{-2mm} \\
                 & Date        & Seeing (\arcsec)   & Exptime (s) & Airmass   & \ionjb{Ca}{2} K          & \ionjb{Ca}{2} H  &  $\overline{V + R}$    & \ha  & $\overline{A + B}$ \\
\vspace{-2mm} \\
\transiti         & 2006/09/07  &   $0.85 \pm 0.22$  & 600 - 900   & 2.1 - 1.6 &21.4                    & 28.0            & 28.3                  & 91.0         & 162.4              \\
\transitii         & 2007/07/19  &   $0.68 \pm 0.11$ & 300         & 2.4 - 1.6 &14.6                    & 18.6            & 19.8                  & 57.8         & 106.1              \\
\transitiii\	 & 2007/08/28  &   $1.17 \pm 0.51$  & 300         & 2.3 - 1.6 &12.5                    & 16.5            & 16.9                  & 52.8         & 94.7               \\ 
\hline
\end{tabular}
\end{center}
\end{table*}

Ground-based transmission spectroscopy of transiting exoplanets is notoriously difficult owing to the small signal and variable nature of the Earth's atmosphere. When spectroscopy is employed, variable seeing changes the slit and hence grating illumination, leading to systematic changes in the recorded spectrum. These are generally second order effects, but are crucial when searching for precision radial velocity signatures or subtle changes in the spectrum at levels of order a few per cent or less. Spectrophotometry has proven to be one successful method enabling multi-wavelength transit radius measurements in the atmospheres of close orbiting planets. For example, \cite{sing12xo2b} used a wide 5\arcsec\ long slit to carry out differential spectrophotometry of the planet hosting star, XO-2A, and its planet, XO-2b, yielding a 5.2-$\sigma$ detection of Na. 

Extending ground-based characterisation to higher resolution is clearly desirable if we wish to detect individual transitions and measure abundances in detail. \'{E}chelle spectroscopy offers the wavelength range needed, but with many optical components, including a grating and a cross-disperser, this presents greater challenges, especially when narrow slit widths of typically $\lesssim 1$\arcsec\ are commonly employed. Fibre fed instruments offer the best internal stability since the light is largely scrambled by the fibre and hence the illumination of the slit on exit from the fibre is less sensitive to seeing changes and guiding errors. 
By normalising each observation to nearby continuum regions, excess absorption in individual lines can be probed. This was first applied specifically to \hdone\ by \cite{redfield08} to detect excess Na absorption during transit and more recently using data from the High Accuracy Radial Velocity Planet Searcher (\harps) by \cite{wyttenbach15}. \cite{louden15hd189733} have also applied a model that accounts for the   
differences in the Doppler shift of absorption from opposite sides of the planet. The planetary absorption line profiles were thus modelled as a function of time during the transit to infer an equatorial eastward jet in the planet's atmosphere.

\cite{fossati13wasp12} hypothesised that extrinsic absorbing gas local to the WASP-12 system and arising from WASP-12b is responsible for the anomalously low flux in the \cahk cores of WASP-12. If gas is uniformly distributed around the star, we would expect the absorption to be constant with time. \cite{fossati10wasp12metals} and \cite{haswell12} showed that there is more near ultraviolet absorption around the phases of transit, so the absorbing gas in the WASP-12 system is densest there. Because WASP-12 is faint, and there is very little flux in the line cores \citep{fossati13wasp12}, we decided to study \hdone to search for time variable \cahk absorption. \hd is a K1.5V star with a relatively small radius of \hbox{$R_* =$} \hbox{$0.756 \pm 0.018$ \rsun} \citep{torres08}. Hence a large transit signature for a given planet radius is expected compared with a star such as WASP-12b with an estimated radius of \hbox{$R_* =$} \hbox{1.6 \rsun}. Conversely, the pressure scale height of $H_{eq} = 201$ km is not particularly high compared with \hdtwo, with $H_{eq} = 553$ km \citep{sing16nature}. 

In this paper, we examine \cahk and \ha absorption during the transit of \hdone. We also re-examine \nad absorption following a recent analysis by \cite{wyttenbach15} and also use the \ionjb{Ca}{1} 6122 \AA\ line as a control. The archival \harps\ data exhibit clear evidence of excess absorption during transit, thanks largely to the fact that the star is active and hence exhibits \cahk cores with sufficient flux to reliably detect the transit. In \S \ref{section:analysis_hd189733}, we discuss the activity and chromospheric variability of \hbox{HD 189733}. The observations and the methods used to derive the transit light curves is presented \S \ref{section:observation}. We present the transit light curves in \S \ref{section:results_lightcurves} and the residual line profiles during transit in \S \ref{section:results_transmission}. We discuss the possible origin of the excess absorption during transit in light of our finding that the absorption is both variable and more sharply defined in the stellar reference frame than the planet reference frame \S \ref{section:discussion}. A summary and concluding remarks are found in \S \ref{section:conclusion}.

\section{Chromospheric and photospheric variability of HD 189733}
\label{section:analysis_hd189733}

The relatively active nature of \hbox{HD 189733} was recognised in the discovery paper by \cite{bouchy05hd189733}, following measurements of the Mount Wilson S index by \cite{wright04activity}. During the observations studied here, the S index \citep{vaughan78}, varies in the range 0.461 to 0.508, with associated $R'_{HK}$ \citep{noyes84} values in the range $-4.55$ $\leq$ log $R'_{HK} \leq -4.50$. For comparison, the Sun typically shows average values of $-5.0$ $\leq$ log $R'_{HK} \leq -4.85$ for solar minimum and solar maximum periods \citep{lagrange10spots}.

Active stars also exhibit greater chromospheric variability than less active stars, as seen when comparing S indices measured in the same stars at different epochs \citep{wright04activity,jenkins06activity}. Variability may result from stellar magnetic cycles, rotational modulation or shorter lived transients such as flares. \cite{shkolnik03hd179949} found evidence for synchronous enhancement of \cahk in \hbox{HD 179949} due to its non-transiting hot Jupiter. In contrast, modulation only at the rotation period of \hbox{HD 189733} was found by \citep{shkolnik08onoff}. Similarly, \cite{boisse08hd189733} found periodicity in the \cahk lines close to the stellar rotation period of \hbox{HD 189733}.

The chromospheric activity variability complicates both radial velocity measurements and transit measurements. \cite{boisse08hd189733} used correlations between chromospheric indicators and radial velocity measurements to improve their orbital solution, and hence the mass estimate of \hdone. Since photospheric distortions from associated cool starspots are the main cause of astrophysical noise in high precision stellar radial velocities, their effects must also be accounted for, particularly where high S/N ratio space-based observations that are free from atmospheric systematics are available. Both \cite{pont08hd189733} and \cite{sing11hd189733} have included spot modelling analysis in precision transmission spectroscopy with Hubble Space Telescope observations of \hdone, to obtain improved estimates of wavelength-dependent transit depths.

In addition to the problems introduced by chromospheric variability, throughput efficiency typically declines in spectrometers at blue wavelengths. For \harps\ the efficiencies at \hbox{3800 \AA}~and \hbox{4000 \AA}~are respectively 0.24 and \hbox{0.47} of its maximum throughput of 5.7 per cent\footnote{http://www.eso.org/sci/facilities/lasilla/instruments/ \hbox{\hspace{2mm}harps/doc.html}}. The S/N ratios achievable in the cores of the strong \cahk lines are thus generally low as the typically low throughput of the instrument is exacerbated by the deep photospheric absorption in these features. Hence, stars exhibiting moderate activity, and thus appreciable line core emission flux, are likely to offer the best chances of detecting a planetary transit in \cahk, providing that the star remains relatively quiet during and around transit. A recent study of \hdone by \cite{czesla15hd189733}, utilising UVES/VLT observations has investigated the limb-darkening effect during transit in the Na doublet and \cahk lines, but the cores of the lines, particularly \cahk, were affected by a flaring event.

\section{Observations and differential spectrophotometry}
\label{section:observation}

\hdone\ has been observed with \harps\ during a number of transit events\footnote{This work is based on observations collected at the European Organisation for Astronomical Research in the Southern Hemisphere under ESO programmes 072.C-0488(E), 079.C-0127(A) and 079.C-0828(A). The data are available at https://archive.eso.org.}. 
{ The \harps Data Reduction Software ({\sc drs}) does not apply background subtraction of scattered light. We find that after bias subtraction, the typical line core fluxes for \ionjb{Ca}{2} H, \ionjb{Na}{1} D1 and \ha include respective scattered light components of 1.3, 2.9 and 0.3 per cent.} Instead of using the standard { {\sc drs}}, we re-extracted the data using a new extraction pipeline. This is based on the Flat-ratio Optimal eXtraction ({\sc fox}) demonstrated in \cite{zechmeister14}.
Prior to the {\sc fox} extraction, special care is take in performing a detailed background subtraction. This can be important in regions where few counts are obtained, such as the cores of strong lines that contain little flux. That is, all the fibre A and B traces containing the science and simultaneous calibration spectra are set to zero weight and a 2-dimensional 3rd-degree polynomial fit is done on both HARPS chips to remove the smooth component of the background. This still leaves small residuals, especially on the edges of the detector and regions contaminated by secondary ghost spectra produced by HARPS. To mitigate these, each pixel column (cross-dispersion, science and calibration spectra still blocked) is also fitted to a parabola and subtracted. 

{ Since the fraction of scattered light is greatest in the line cores, any further flux deficit during transit will be more significant in the case where the scattered light is removed during extraction. This will lead to greater transit depth measurements. The transits we measure in \S \ref{section:results_lightcurves} are $\sim 2.5$ per cent deeper for \cahk and \nad using our extraction compared with the {\sc drs} spectra. We note that the significance of this finding is nevertheless at a level lower than our measurement uncertainties.}

We use the same naming scheme as \cite{wyttenbach15} who studied transits of the Na line using the same data. \transiti data were obtained on 2006/09/07 with exposure times between 600 secs and 900 secs. \transitii and 3 data were taken on 2007/07/19 and 2007/08/28 respectively, with 300 sec exposures. In Table 1, we summarise the observations, and give S/N ratios for the \cahk, and \ha regions considered in this paper. Column 4 of Table \ref{table1} shows that the observations were made at high airmass owing to the northerly declination ($\delta = 22.6$\degs). The seeing recorded in the observation FITS headers indicates variable conditions at each epoch. Column 3 of Table \ref{table1} shows that the best seeing was present during \transitii (0.68\arcsec), which also afforded the most stable conditions. The seeing during \transitiii, with a mean of 1.17\arcsec, was substantially worse. Since the \harps\ fibres are 1\arcsec\ on sky, we expect the most efficient and stable spectra to have been observed during \transitii.

Atmospheric variability generally precludes precision flux measurements with spectrometers fed by a slit of comparable width to the seeing PSF. The accuracy required can nevertheless be achieved by self-calibrating the spectrum using simultaneous differential measurements. Using this method, \cite{charbonneau02atmos} made differential measurements of the Na lines by comparing in-transit (IT) and out-of-transit (OOT) fluxes centred on Na with those in neighbouring continuum bands. This is a doubly differential method: the difference between IT and OOT and the difference between the line and continuum is considered. It can only be used to reliably infer absorption from the planet atmosphere if the stellar spectrum is completely constant with time.
This approach has been adopted by a number of authors including \cite{wyttenbach15}, who applied it to the \harps\ observations we use here. In this way, the signature of continuum light blocked by the opaque planet is removed. With no wavelength-dependent signal (i.e. additional line absorption relative to the continuum) we thus expect no sign of transit, while excess absorption during IT phases can be attributed to the presence of a line in the planetary atmosphere. 

\subsection{Transit light curve measurement of Ca {\sc ii} H \& K lines}
\label{section:analysis_transit_ca}

The differential transit light curve method is analogous to the Mount Wilson method of measuring \cahk: the S index combines both H \& K line core fluxes to improve the S/N ratio (see Table \ref{table1}). Here we consider the fractional flux in the \cahk lines relative to the nearby continuum bands. The unweighted line core flux,  $\mathscr{U}_{HK}(t)$, for an observation at time, $t$, can be written as

\begin{equation}
 \mathscr{U}_{HK}(t) = \frac{\overline{K}(t) + \overline{H}(t)}{\overline{V}(t) + \overline{R}(t)}
 \protect\label{equation:simple_flux}
\end{equation}

where $\overline{K}(t)$ and $\overline{H}(t)$ are the mean fluxes in passbands centred on the \ionjb{Ca}{2} K and \ionjb{Ca}{2} H lines respectively. Similarly, the blue continuum band flux is $\overline{{V}}(t)$ and the red continuum band flux is $\overline{{R}}(t)$. We stress that the line fluxes are {\em not} the same as the Mount Wilson fluxes which are weighted with a triangular passband \citep{vaughan78} of width 1.09 \AA. The exact width of the H \& K passbands are however important, and should include the wavelength range over which any IT vs OOT transit variability is seen. Although the overall flux in H is lower than in K, the S/N ratio in H is higher, as indicated in Table \ref{table1}. We therefore optimally add the mean fluxes according to their associated variances, $\overline{\sigma_H}^2$ and $\overline{\sigma_K}^2$, which are calculated from the individual pixel variances. Hence, for our analysis by replacing the numerator, $\overline{K}(t) + \overline{H}(t)$ in Equation \ref{equation:simple_flux} with

\begin{equation}
 \overline{{HK}}(t) = \frac{\overline{K}(t)/\overline{{\sigma_{K}}}^2(t) + \overline{H}(t)/\overline{\sigma_{H}}^2(t)} {1/\overline{\sigma_K}^2(t) + 1/\overline{\sigma_H}^2(t)}
\protect\label{equation:weighted_HK}
\end{equation}

\noindent
We can thus write the combined, weighted \cahk line core flux, as

\begin{equation}
 \mathscr{F}_{HK}(t) = \frac{ \overline{{HK}}(t)}{\overline{V}(t) + \overline{R}(t)}
 \protect\label{equation:HK_flux}
\end{equation}

\noindent
The $V$ and $R$ band fluxes are not optimally added, as this may introduce airmass-dependent normalisation systematics. This is especially important as altitude dependent atmospheric refraction is a steep function at blue wavelengths. 

The mean S/N ratios are listed in Table \ref{table1} for the \ionjb{Ca}{2} K,  H and the combined V and R bands. The line passbands used for the S/N ratio statistics were 0.8 \AA\ wide, centred on each line. For the \ionjb{Ca}{2} lines, we adopted the same 20 \AA\ wide V and R continuum passbands as used for Mount Wilson S index measurements. The V passband spans 3891.07\,-\,3911.07 \AA\ and the R band spans 3991.07\,-\,4011.07 \AA.

\begin{figure*}
\begin{center}
\begin{tabular}{cc}
\includegraphics[scale=0.34,angle=270]{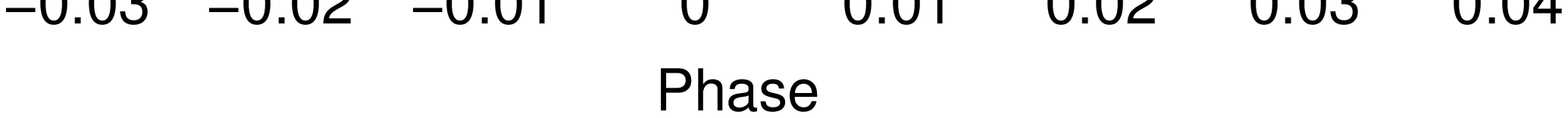} & \hspace{-4mm}
\includegraphics[scale=0.34,angle=270]{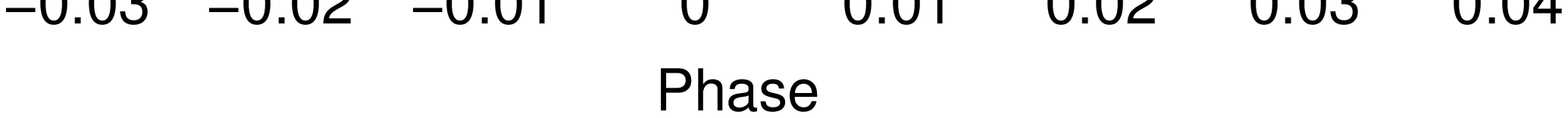} \\
\includegraphics[scale=0.34,angle=270]{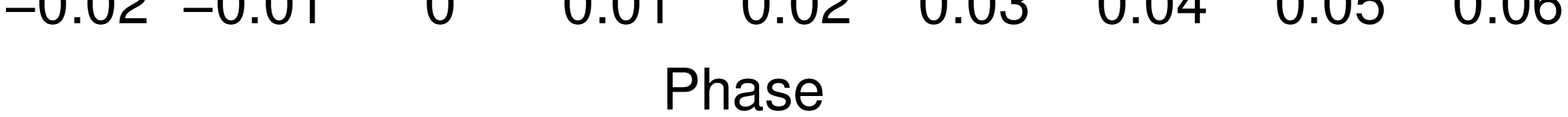} & \hspace{-4mm}
\includegraphics[scale=0.34,angle=270]{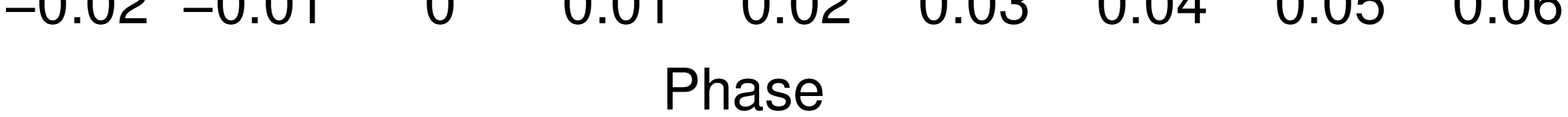} \\
\includegraphics[scale=0.34,angle=270]{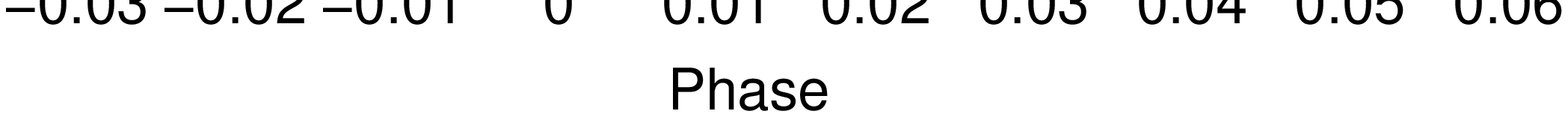} & \hspace{-4mm}
\includegraphics[scale=0.34,angle=270]{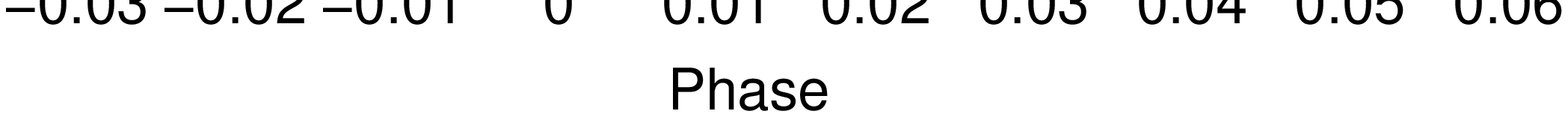} \\
\end{tabular}
\end{center}
\caption{Differential transit fluxes for \cahk and \ha with the signature of the continuum light blocked by the opaque planet removed. The panels show \transiti (a \& b) and \transitii (c \& d) and \transitiii (e \& f). For \cahk, line core fluxes are calculated in a combined 0.4 \AA\ (H) and 0.6 \AA\ (K) passband (a, c \& e). For \ha, line fluxes are calculated in a 0.7 \AA\ passband. The $\phi_1$ to $\phi_2$ and $\phi_3$ to $\phi_4$ partial transit phases are shown by the darker grey shaded regions, while the full transit phases $\phi_2$ to $\phi_3$ are denoted by the lighter grey regions. The transit depth is determined from the mean difference between the OOT flux, where $\phi < \phi_1$ and $\phi > \phi_4$ (white regions) and the IT flux, where $\phi_2 < \phi < \phi_3$. The transit outlines are intended to guide the eye and are not transit model fits that include geometric and limb-darkening effects. The \transiti data have been normalised using all OOT observations (solid/black lines) and for \ha, excluding the first 2 observations after $\phi_4$ (dashed/grey line), which are possibly associated with a flare. Similarly, for the \transitii \cahk light curves (c), the transit outlines are shown by the solid/black line while the dashed/grey line shows the transit without the outlying observation at $\phi = -0.0039$. \transitiii light curves are affected by a rise in emission, characteristic of a flare. Tentative transit depths calculated using the partial transit regions for \transitiii are indicated and shown by dashed lines in panels e \& f (see \S \ref{section:results_night3} for details).
\label{fig:FluxHK}}
\end{figure*}

\begin{figure*}
\begin{center}
\includegraphics[scale=0.65,angle=270]{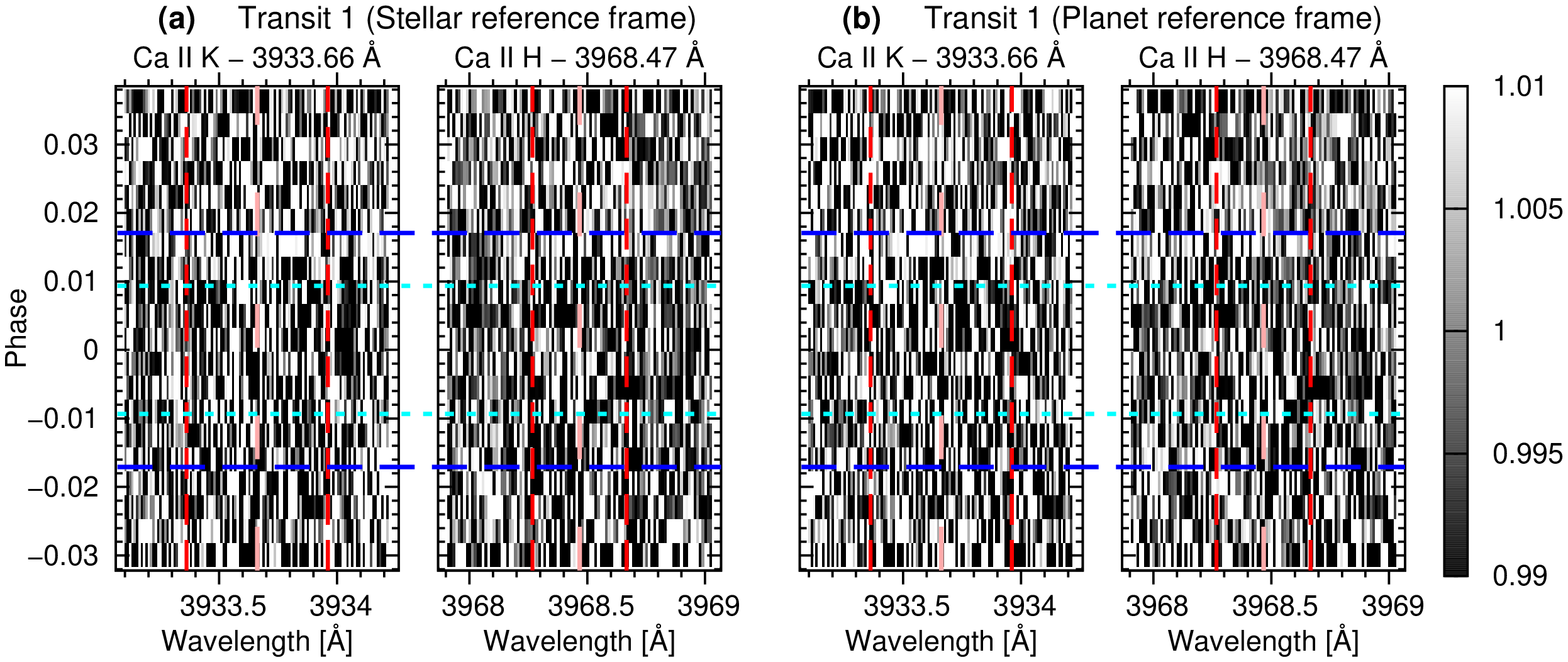} \\
\vspace{3mm}
\includegraphics[scale=0.65,angle=270]{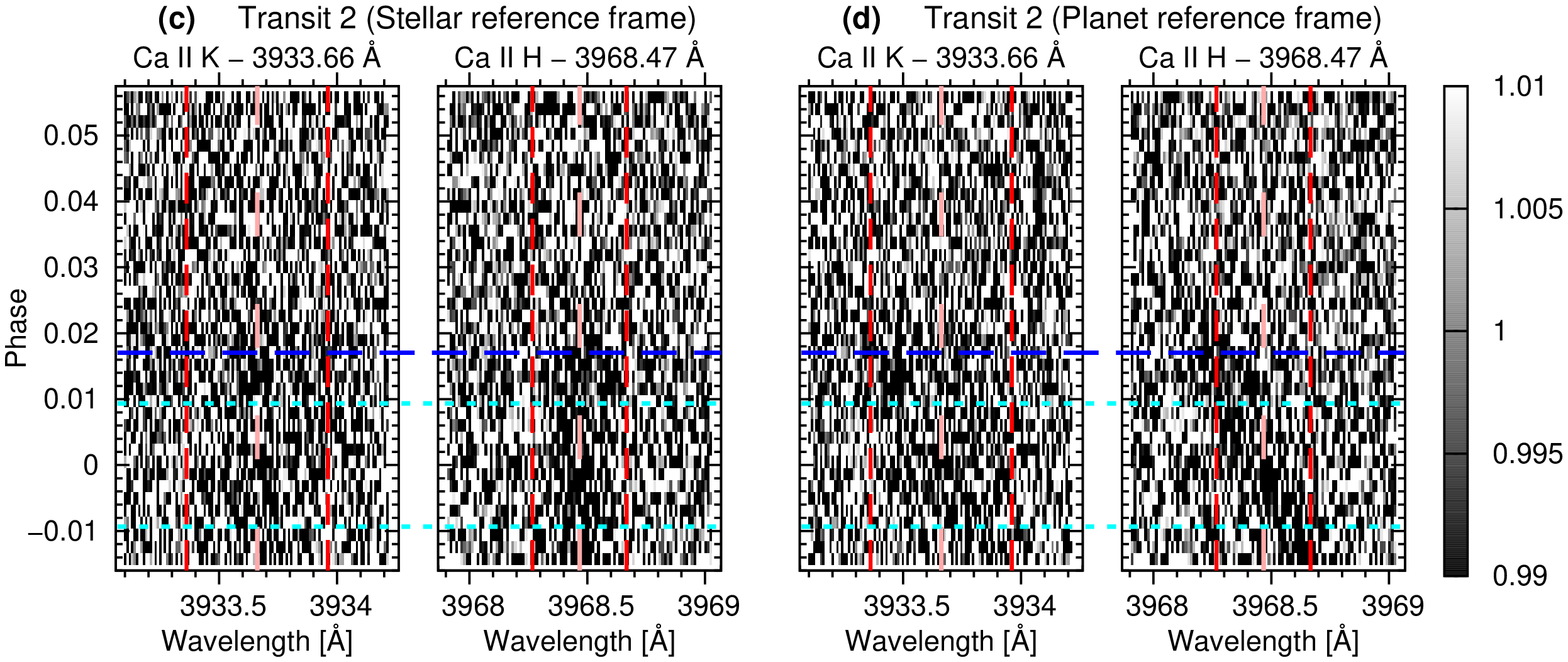} \\
\vspace{3mm}
\includegraphics[scale=0.65,angle=270]{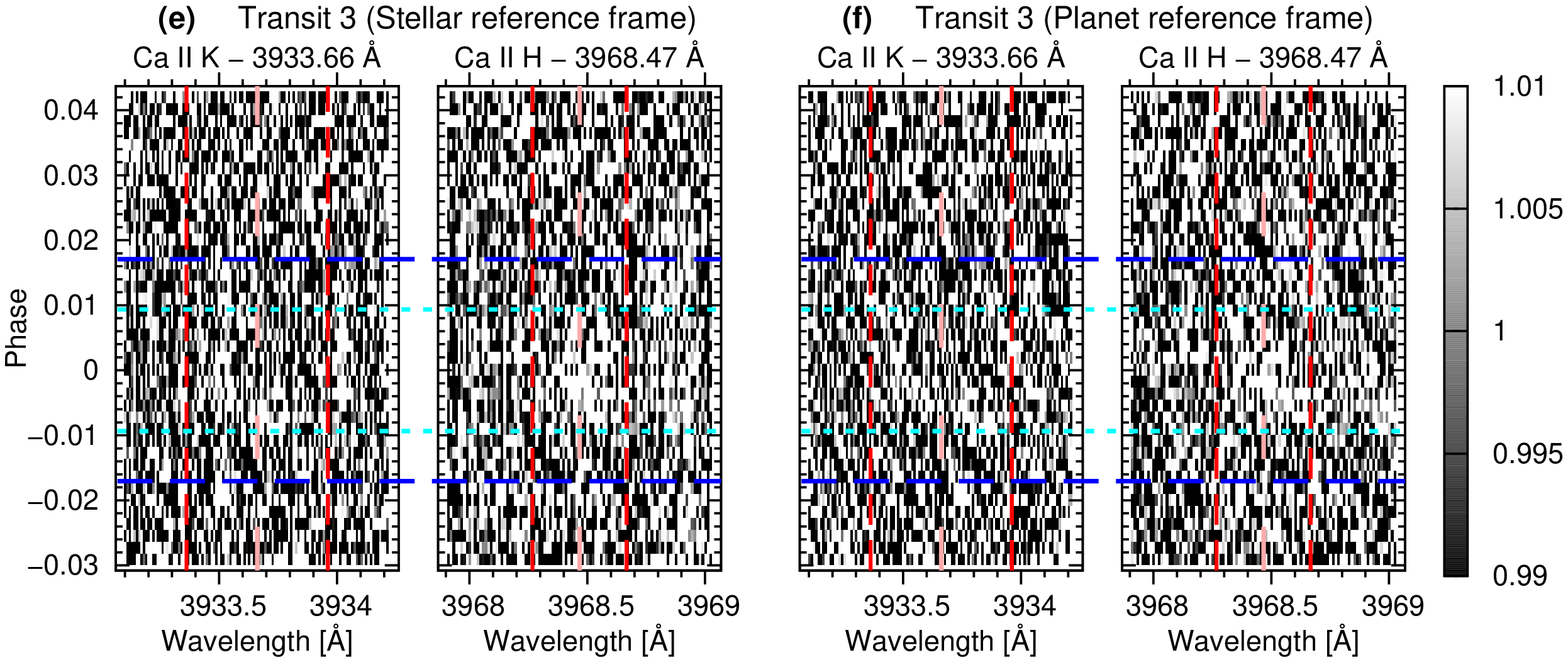} \\
\caption{Normalised residual \cahk time series for each transit of HD 189733b: \transiti (panels a \& b), \transitii (panels c \& d) and \transitiii (panels e \& f). The left hand panels (a, c \& e) are the stellar reference frame time series, while the right hand panels (b, d \& f) are time series shifted into the planetary reference frame. The rest wavelength of each line is indicated by a vertical, long-dashed/pink line. The vertical dashed/red lines indicate the central $\pm 0.3$ \AA\ and $\pm 0.2$ \AA\ ($\Delta\lambda\ =\ 0.6\ \&\ 0.4$) regions bracketing the emission cores of \ionjb{Ca}{2} K and \ionjb{Ca}{2} H respectively. The first and fourth contact points of the transits, $\phi_1$ and $\phi_4$, are indicated by the dashed/blue lines. $\phi_2$ and $\phi_3$ are marked by the dotted/cyan lines. \label{fig:dynHK}}
\end{center}
\end{figure*}

\subsection{Transit light curve measurement of \ha}
\label{section:analysis_transit_ha}

For \ha, we use Equation \ref{equation:simple_flux} with one line only, such that 

\begin{equation}
\mathscr{F}_{H_{\alpha}}(t) = \frac{\overline{{H_\alpha}}(t)}{{\overline{A}(t) + \overline{B}(t)}}
\protect\label{equation:halpha_flux}
\end{equation}

\noindent
The S/N ratio statistics for the \ha and the combined A and B bands (\hbox{$6422 - 6448$ \AA\ and} \hbox{$6640 - 6660$ \AA}\ respectively) are tabulated in columns 8 \& 9 of Table \ref{table1}. Although the adopted continuum regions were chosen to ensure strong telluric H$_2$O lines are excluded, very weak lines of $\lesssim 1$ per cent of the normalised continuum are present in these passbands. For the \ha line itself, depending on adopted width of the band, two significant telluric lines contaminate the profile. The wavelengths of these lines are \hbox{6562.45 \AA}\ and \hbox{6563.51 \AA}\ with respective depths of \hbox{$\sim2$}\ per cent and \hbox{$\sim7.5$}\ per cent relative to the normalised continuum. We used the Line By Line Radiative Transfer Model ({\sc lblrtm}) code \citep{clough92,clough05} to obtain model telluric spectra using temperature and pressure soundings from the Air Resources Laboratory\footnote{http:\/\/ready.arl.noaa.gov\/READYamet.php}, as outlined in \S 4.3 of \cite{barnes12rops}. To account for any mismatches in radial velocity, spectral resolution and line strength (which depends on the model H$_2$O column density), the telluric spectra were then scaled to the observed spectra using the technique outlined in Appendix A of \cite{cameron02upsand}. Dividing each spectrum by the scaled telluric model appropriate to that observation results in effective removal of the telluric lines. These spectra are then used to perform the subsequent analysis using Equation \ref{equation:halpha_flux} above. 

\subsection{Transit light curve measurement of \nad and \ionjb{Ca}{1} 6122 \AA}
\label{section:analysis_transit_na_ca1}

We also computed transit light curves for the \nad lines and for \ionjb{Ca}{1} 6122 \AA\ line. As for \cahk, we employed Equation \ref{equation:HK_flux} to calculate the \nad fluxes, while for \ionjb{Ca}{1} 6122 \AA, Equation \ref{equation:halpha_flux} was used.

\subsection{Transit depth}
\protect\label{section:transit_depth}

The transit light curves obtained from applying Equations \ref{equation:HK_flux} \& \ref{equation:halpha_flux} are normalised to the OOT flux and then used to determine the transit depth. Since the IT flux difference is measured relative to the OOT flux, which is also non-zero, the measured transit depth depends on the width of the $H$, $K$ and $H\alpha$ passbands. The depth, $t_d$, for a passband, or combined passband width of $\Delta\lambda$ is thus:

\begin{equation}
t_d(\Delta\lambda) = 1 - \frac{\overline{\mathscr{F}}_{IT}}{\overline{\mathscr{F}}_{OOT}}
\protect\label{equation:transit_depth}
\end{equation}

\noindent
where $\overline{\mathscr{F}}_{IT}$ and $\overline{\mathscr{F}}_{OOT}$ are the mean IT and OOT fluxes calculated in the passband of width $\Delta\lambda$ over the appropriate phases. We choose to define the equation this way so that $t_d$ is a positive value if excess absorption is detected. { For the transit light curve analysis, we define IT phases as those that lie between contact points 2 and 3 \citep{haswell2010book}}. \cite{winn07hd189733b} measured { these} contact points as $\phi_3 = -\phi_2 =  0.00946$ and $\phi_4 = -\phi_1 = 0.01716$. Simultaneous modelling of both photometric and spectroscopic data has yielded $\phi_4 = 0.01696$ \citep{triaud09hd189733} from which we estimate $\phi_3 =  0.00935$ by scaling the \cite{winn07hd189733b} $\phi_3/\phi_4$ ratio. Using the parameters given by \cite{agol10hd189733} and the radius of \cite{sing11hd189733} defined in the 3700\,-\,4200 \AA\ band (i.e. the band containing \cahk), the analytical equations defining the total transit duration, $t_T$, and full transit duration, $t_F$ \citep{seager03transit}, yield \hbox{$\phi_3 =  0.00935$} and \hbox{$\phi_4 = 0.01707$}. We adopt these values at all wavelengths, but note that for \ha, where the radius of \hdone\ is slightly smaller (i.e. assuming the 6500\,-\,7000 \AA\ band radius given by \citealt{pont08hd189733}), the contact points may be very slightly modified \hbox{$\phi_3 =  0.00936$} and  \hbox{$\phi_4 =  0.01706$}.

\begin{table*}
\begin{center}
\caption{\transitii light curve depth measurements for \cahk combined and for \ha. Bandwidths over which flux is measured are given on columns 1, 2, 3 \& 6. The transit depths, $t_d$, are tabulated in columns 4, 5 \& 7 and include the propagated formal uncertainties. The related significance of $t_d$ is shown in brackets. The transit depths for \cahk excluding the outlying observation at phase $\phi = -0.0039$ are listed in column 5. \label{table2}}
\begin{tabular}{cccllcl}
\hline
\multicolumn{5}{c}{\ionjb{Ca}{2} H \& K}   &                                                                  \multicolumn{2}{c}{\ha} \\
\hline
$\Delta\lambda_H$ & $\Delta\lambda_K$ & $\Delta\lambda_{H+K}$  & ~~~~~~~~~~~~~~~~$t_d$              & ~~~~~~~~~~~~~~~~$t_d$                    & $\Delta\lambda$  &  ~~~~~~~~~~~~~~~~$t_d$              \\
\multicolumn{3}{c}{[\AA]}               &                                   & (excluding $\phi = -0.0039$)                  &    [\AA]                   &                                    \\
\hline                                    
0.3 & 0.3 & 0.6                         & 0.0265 $\pm$ 0.0031 (8.5$\sigma$) & 0.0276 $\pm$ 0.0031 (9.0$\sigma$)  & 0.6                   &  { 0.0124 $\pm$ 0.0012 (10.0$\sigma$)}  \\
0.4 & 0.4 & 0.8                         & 0.0223 $\pm$ 0.0024 (9.4$\sigma$) & 0.0231 $\pm$ 0.0023 (9.9$\sigma$)  & 0.7                   &  { 0.0122 $\pm$ 0.0012 (10.2$\sigma$)}  \\
0.4 & 0.6 & 1.0                         & 0.0215 $\pm$ 0.0024 (9.1$\sigma$) & 0.0226 $\pm$ 0.0022 (10.1$\sigma$) & 0.8                   &  { 0.0111 $\pm$ 0.0011 (9.8$\sigma$) } \\
0.6 & 0.6 & 1.2                         & 0.0180 $\pm$ 0.0020 (8.6$\sigma$) & 0.0188 $\pm$ 0.0021 (9.2$\sigma$)  & 1.0                   &  { 0.0103 $\pm$ 0.0010 (9.9$\sigma$)}  \\
0.8 & 0.8 & 1.6                         & 0.0161 $\pm$ 0.0022 (8.5$\sigma$) & 0.0172 $\pm$ 0.0020 (7.3$\sigma$)  & 1.3                   &  { 0.0079 $\pm$ 0.0011 (7.2$\sigma$) } \\
\hline
\end{tabular}
\end{center}
\end{table*}

\section{Transit light curves}
\protect\label{section:results_lightcurves}

In Fig. \ref{fig:FluxHK}, the normalised transit light curves are shown for Transits 1\,-\,3. The passband width, $\Delta\lambda$, was adjusted to maximise the confidence in the measured value of $t_d$. The procedure is described in \S \ref{section:results_night2} below { and removes the transit of the opaque planet, so these curves are additional signal in the line.} The contact phases { for the \hdone transit \citep{winn07hd189733b}, $\phi_1$\,-\,$\phi_4$,} are indicated by the shaded regions in Fig. \ref{fig:FluxHK}. The light curves are all normalised to unity outside transit. With 300 sec exposures, the sampling cadence during \transitii shows the clearest evidence of excess absorption. However, owing to the short interval for which \hdone\ is visible, the transit is not complete. The 600\,-\,900 sec exposures during \transiti yielded the highest S/N ratios of all three transits, but at the cost of temporal resolution. \transiti also shows evidence for additional absorption in both light curves during transit, but the OOT levels are less reliably determined as there are only 9 OOT observations. Both Ca and \ha indicate flaring activity during \transitiii, with a sharp rise in flux between $\phi_1$ and $\phi_2$ followed by a gradual decay. The seeing during \transitiii was also the least stable of the three sets of observations.

\begin{figure*}
\begin{center}
\begin{tabular}{cc}
\includegraphics[scale=0.65,angle=270]{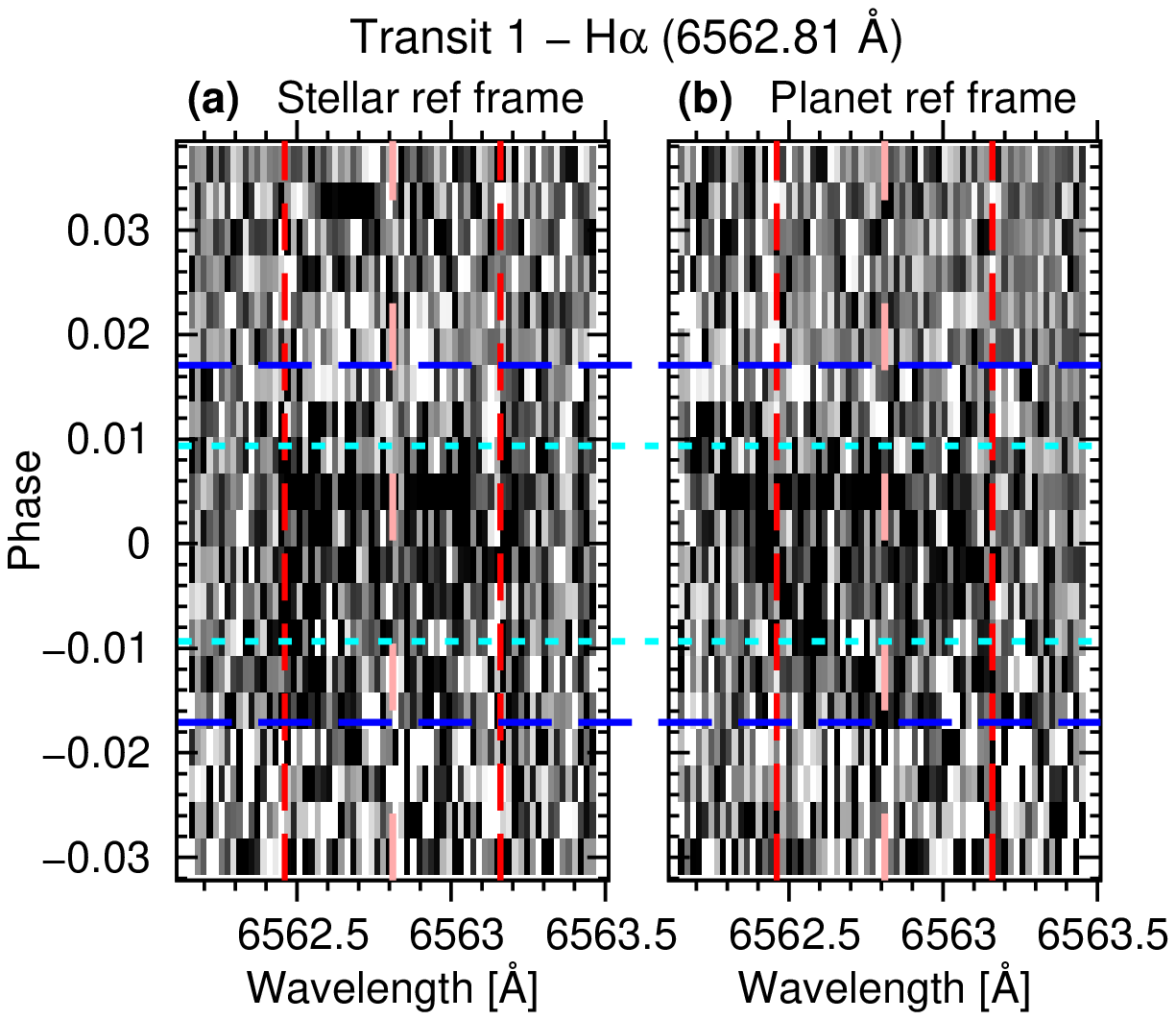} &
\hspace{-6mm}
\includegraphics[scale=0.65,angle=270]{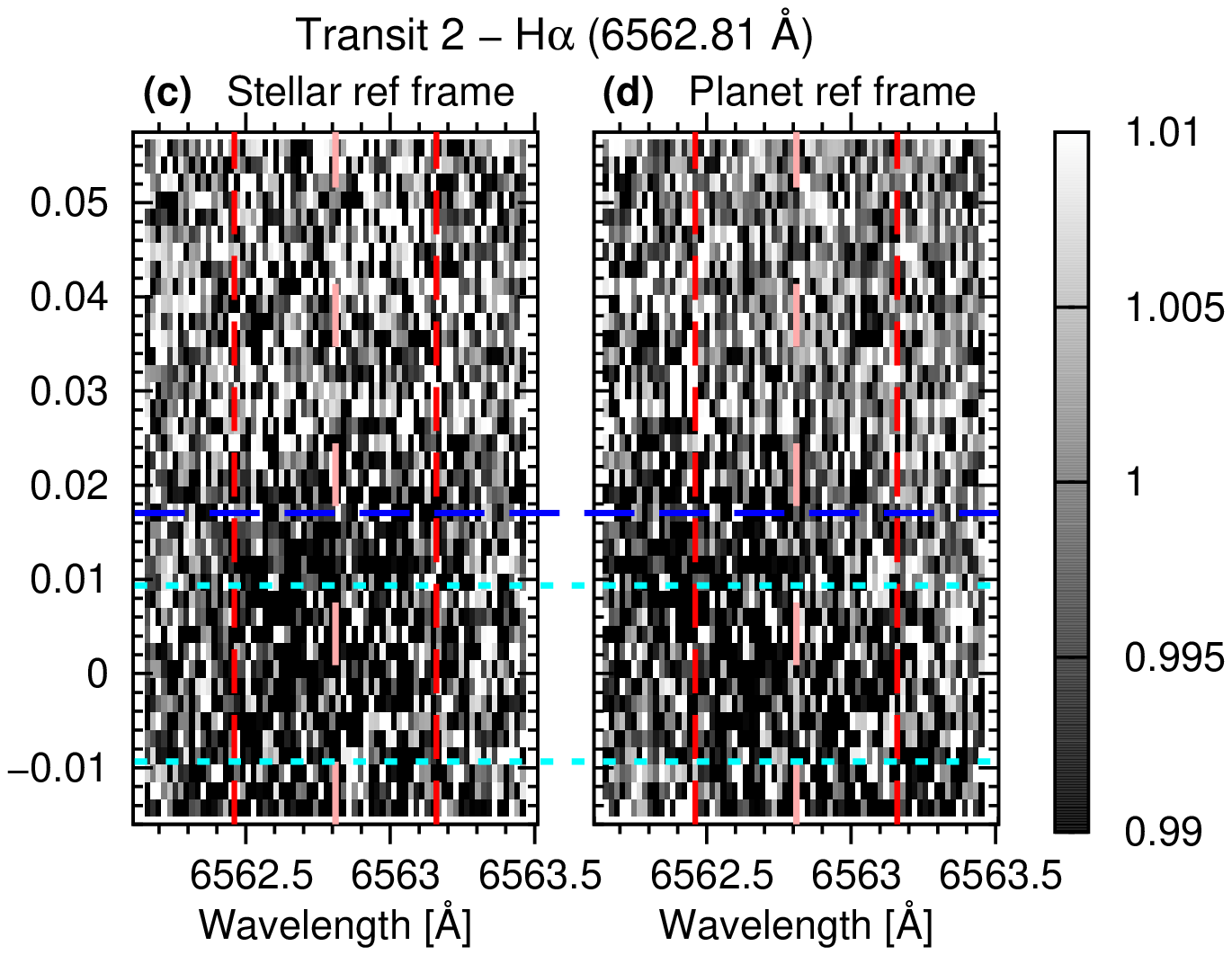} \\
\end{tabular}
\begin{tabular}{c}
\includegraphics[scale=0.65,angle=270]{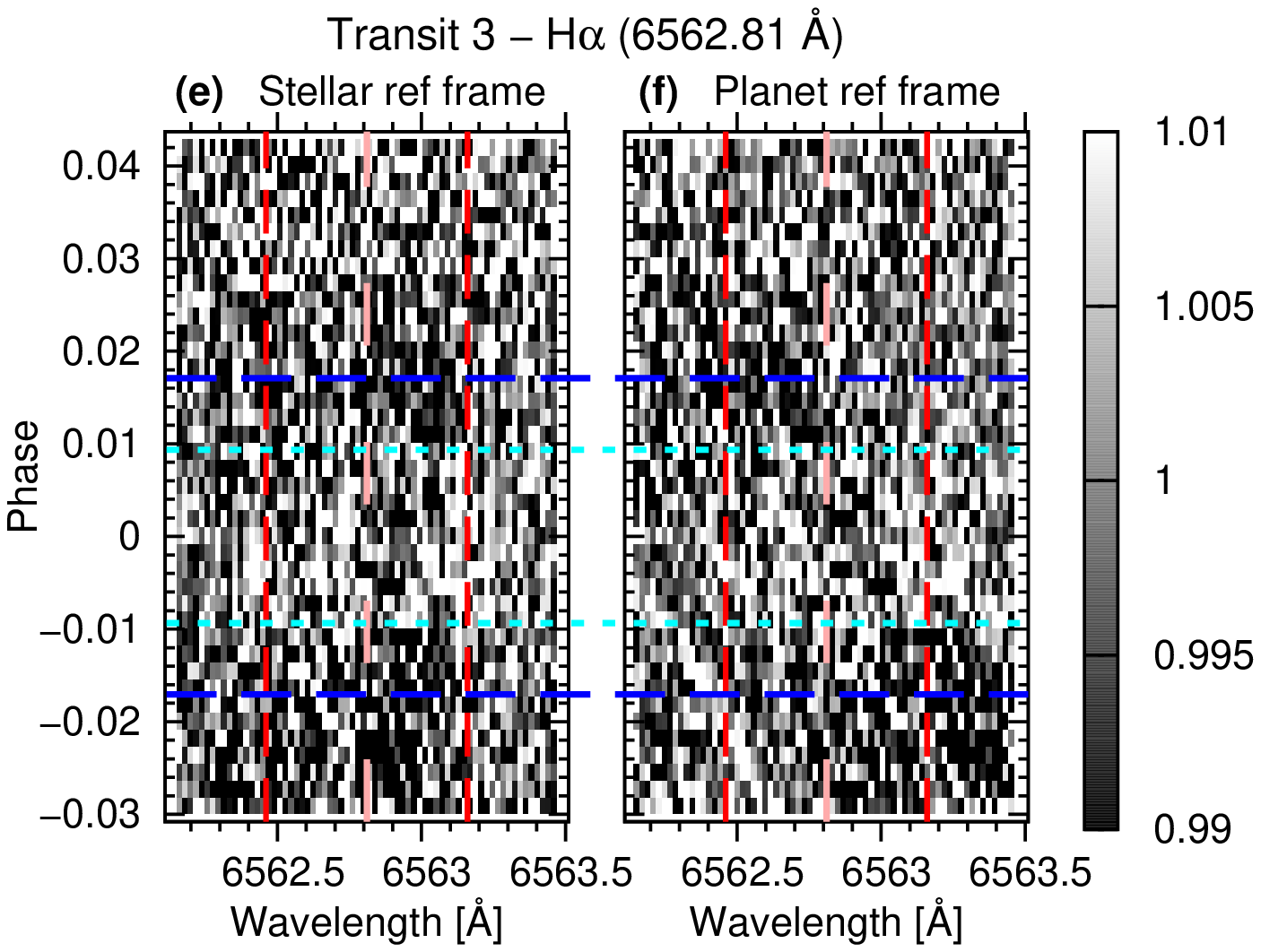} \\
\end{tabular}
\end{center}
\caption{As for Fig. \ref{fig:dynHK} with the \ha (6562.81 \AA) line. The vertical dashed/red lines indicate the central $\pm 0.35$ \AA\ ($\Delta\lambda\ =\ 0.7$\ \AA) for which the excess absorption during transit is optimised. \label{fig:dynHa}}
\end{figure*}

The normalised residual time series spectra of the \cahk and \ha lines are plotted in Fig. \ref{fig:dynHK} and Fig. \ref{fig:dynHa} respectively (panels a, c \& e). The time series in panels a, c \& e were created by firstly dividing each spectrum by the mean continuum passband values defined by the denominator in Equations \ref{equation:HK_flux} \& \ref{equation:halpha_flux} i.e. the \cahk lines were divided by $\overline{V}(t)+\overline{R}(t)$ and the \ha spectra were divided by $\overline{A}(t)+\overline{B}(t)$. From the continuum-normalised spectra, the mean OOT (as defined above in \S \ref{section:transit_depth}) spectrum for each region was calculated and used to normalise each spectrum in turn.
{ This divides out any non-variable spectral features: Figs. \ref{fig:dynHK} \& \ref{fig:dynHa} are trailed spectra showing the variations from the mean. Contact phases are indicated by the horizontal dashed and dotted lines (see Fig. \ref{fig:dynHK} caption for details). For Transits 1 \& 2, the \ha time series in Fig. \ref{fig:dynHa} (a, c \& e) clearly show evidence for additional absorption during the IT period.} The excess absorption in \cahk is only clearly seen during \transitii in Fig. \ref{fig:dynHK}. Both \cahk, and \ha show the emission attributable to stellar activity during \transitiii, which is revealed in the transit light curves in Fig. \ref{fig:FluxHK}, and which precludes measurement of any excess absorption. We attribute this to a flaring event that takes place near the start of \transitiii.

In \S \ref{section:results_night2} below, we initially focus our analysis on the 39 \transitii observations, which show the strongest evidence of excess absorption in the \cahk and \ha lines. The 20 observations from \transiti are then considered in a similar manner in \S \ref{section:results_night1} with brief comments on the \transitiii observations in 
\S \ref{section:results_night3}.

\begin{figure*}
\begin{center}
\includegraphics[scale=0.68 ,angle=270]{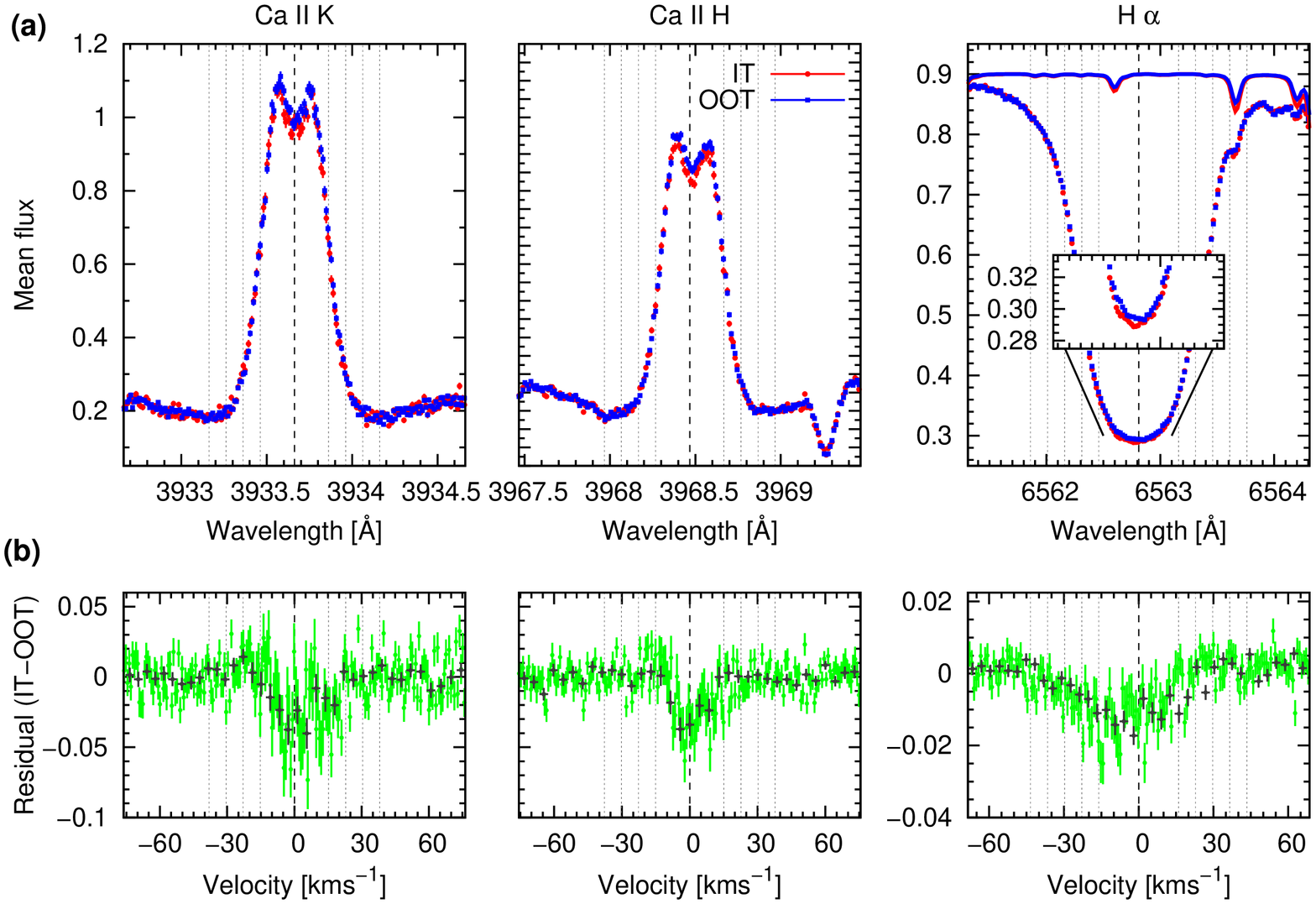} \\
\caption{(a) \transitii averaged \cahk and \ha lines in transit (IT, red circles) and out of transit (OOT, blue squares). All points are plotted with uncertainties. The out of transit profiles were normalised to the mean continuum regions, as described in \S \ref{section:results_night2} before averaging. The inset for \ha is plotted with the same x-scale as the main axis, but with an expanded y-axis to enable the IT/OOT flux difference to be seen. The vertical dashed lines indicate the rest wavelength of each spectral line. For the \cahk plots, the dotted vertical lines show the $\Delta\lambda = 0.3, 0.4, 0.6\ \&\ 0.8$ widths in Table \ref{table2} over which the line fluxes were calculated. Similarly for \ha, the $\Delta\lambda = 0.6, 0.7, 0.8, 1.0\ \&\ 1.3$ regions are indicated. (b) Residual IT - OOT profiles in the reference frame of \hbox{HD 189733}. Unbinned points are shown in green, while the black points are averaged into 0.05 \AA\ and 0.075 \AA\ bins for \ionjb{Ca}{2} and \ha respectively. Bin widths are indicated on the residual plots by the horizontal bars. \label{fig:IT_OOT}}
\end{center}
\end{figure*}

\subsection{\transitii light curves}
\protect\label{section:results_night2}

Fig. \ref{fig:IT_OOT} (a) shows the core regions of \ionjb{Ca}{2} K, H and \ha for IT and OOT spectra. The lines have been corrected to the rest wavelength in air, denoted by the long-dashed vertical lines (the velocity corrections ranged from -6.8 to -6.5 \kms\ during the observations). The telluric lines present in the \ha line, which vary in strength with airmass, have been corrected for using the method outlined in \S \ref{section:analysis_transit_ha}. For \transitii, 11 exposures fall within the $\phi_2$ - $\phi_3$ IT range, while 19 exposures were taken at OOT phases (i.e. taken at phases $> \phi_4$). Looking at the centres of the line profiles, it is clear that there is more absorption in all three lines during IT compared with OOT phases. The residuals are plotted in Fig. \ref{fig:IT_OOT} (b) and show the mean excess absorption in each case, with the wavelength scale replaced by the equivalent velocity in the rest frame of each line. 
{ The width of the additional IT absorption appears to be greater in the emission core of \ionjb{Ca}{2} K compared with \ionjb{Ca}{2} H. \ionjb{Ca}{2} K is the stronger line in stellar spectra because while the \ionjb{Ca}{2} H and \ionjb{Ca}{2} K arise from a transition with the same lower state, the upper states differ. The different degeneracies of the upper two levels results in an enhanced source function of K over H in the upper chromosphere of the star \citep{rauscher06}. Measurement of transit depths must take account of the different widths of any excess absorption by adapting passband widths for flux measurement.}

The transit excess due to Na in the atmosphere of \hdtwo\ was measured by \cite{charbonneau02atmos}, who considered the flux in passbands of three different widths. Subsequent analyses by \cite{redfield08}, \cite{snellen08hd209458} and more recently by \cite{wyttenbach15} using the spectra discussed herein, have also investigated the significance of the detection with different passband widths. The method of flux measurement outlined in \S \ref{section:observation} does not account for the radial velocity of the planet during transit. As noted by \cite{wyttenbach15}, the radial velocity of \hdone\ spans $-15.9$ to $+15.9$ \kms\ in the $\phi_1\,-\,\phi_4$ interval. { Between $\phi_2$ and $\phi_3$}, the range of velocities is $-8.47$ to $+8.47$ \kms. We consider only phases $\phi_2 < \phi < \phi_3$ when measuring the transit depth. At air wavelengths of $\lambda =$ \hbox{3933.66} \AA, \hbox{3968.47 \AA}\ and \hbox{6562.79 \AA}\ for \hbox{\ionjb{Ca}{2} H, K} and \ha \citep{kramida15nist}, absorption from \hdone\ in the $\phi_2$ - $\phi_3$ range will lie within \hbox{$\pm\,0.113$ \AA}\, \hbox{$\pm\,0.114$ \AA}\ and \hbox{$\pm\,0.188$ \AA}\ respectively. Hence minimum filter widths for calculating the line fluxes of \hbox{$\Delta\lambda = 0.226$ \AA}, \hbox{$0.228$ \AA}\ and \hbox{$0.377$ \AA}\ should be used to ensure all absorption co-moving with the planet is included in the IT observation phases.

\begin{figure}
  \begin{center}
    \begin{tabular}{l}
      \hspace{-3mm}
      \includegraphics[scale=0.335,angle=270]{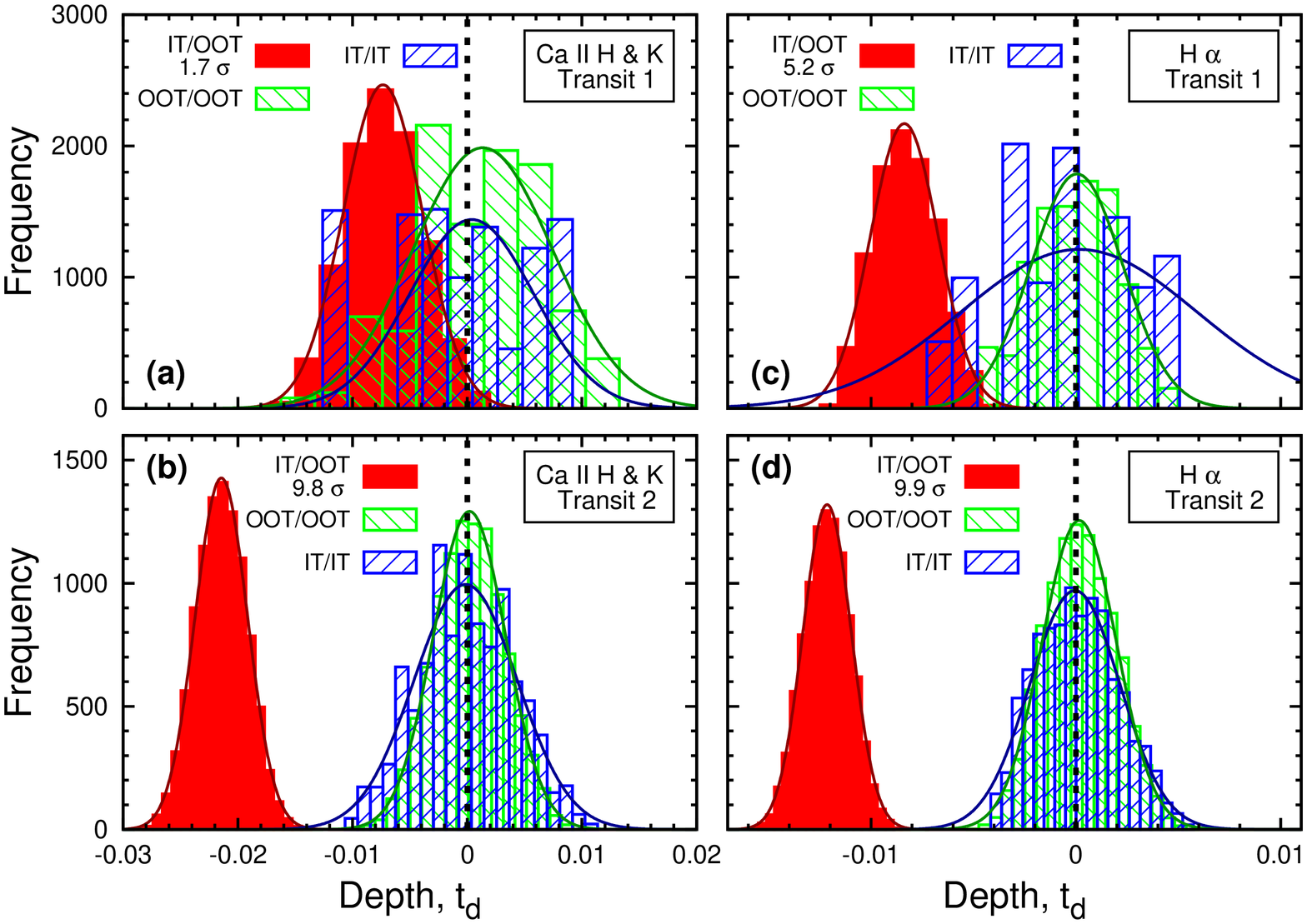} \\
    \end{tabular}
  \end{center}
  \caption{Empirical Monte Carlo analysis of the Transit 1 \& 2 light curves for \cahk and \ha. \label{fig:mc_all}}
\end{figure}

\subsubsection{\cahk}
\protect\label{section:results_night2_ca}
For the \cahk lines, we measured the transit depth in passbands of $\Delta\lambda\ =\ 0.3,\ 0.4,\ 0.6\ \&\ 0.8$ \AA\ for each line (i.e. combined passband widths of $\Delta\lambda\ =\ 0.6,\ 0.8,\ 1.2\ \&\ 1.6$ \AA). These passband limits are marked by the short-dashed lines in Fig. \ref{fig:IT_OOT}. Since the absorption feature in the \ionjb{Ca}{2} K line is broader than the \ionjb{Ca}{2} H line, we also tried a combined passband of $0.4$ \AA\ for H and $0.6$ \AA\ for K, giving a 1.0 \AA\ effective total passband. The results are listed in Table \ref{table2}. The individual passband widths for \ionjb{Ca}{2} H and K are listed in columns 1 \& 2, along with the combined passband, which is simply the sum of columns 1 \& 2. Using equal passbands and all observations to measure $t_d$, the transit depth is detected with greatest significance for the combined $2 \times 0.4$ \AA\ passband, with $t_d = 0.223 \pm 0.0024$. We also investigated removal of the outlying point at $\phi = -0.0039$, which may simply be a statistical outlier. With this point removed, the \cahk transit depth, $t_d$, is detected with greatest significance for the combined passband of 1.0 \AA, which utilises the \ionjb{Ca}{2} H and \ionjb{Ca}{2} K widths of $0.4$ \AA\ and $0.6$ \AA\ respectively. Fig. \ref{fig:IT_OOT} (a) shows that these respective widths correspond to the widths of the core emission reversals as defined by the points either side of the centre of the line at which the second derivative of the flux with respect to wavelength is maximised. With a combined 1.0 \AA\ passband, we detect additional absorption due to \ionjb{Ca}{2} with $t_d = 0.0226 \pm 0.0022$ ($2.26 \pm 0.22$ per cent) at a significance of 10.1$\sigma$. 

\subsubsection{\ha}
\protect\label{section:results_night_ha}
For \ha, the use of single passbands in the range \hbox{$\Delta\lambda =$} \hbox{$0.6\,-\,1.3$ \AA}\ were investigated. Table \ref{table2} shows that excess absorption due to \ha is most significantly detected in a 0.7 \AA\ passband, { with  $t_d = 0.0122 \pm 0.0012$ ($1.22 \pm 0.12$ per cent) and a significance of 10.2$\sigma$}. Despite using only a single line, the { same} significance compared with the excess absorption in \cahk can be attributed to the higher S/N ratios at \ha (see Table \ref{table1}).

\subsection{\transiti light curves}
\protect\label{section:results_night1}

We used the same optimised passbands as for \transitii to obtain \transiti light curves, which also show excess absorption, albeit with lower significance \hbox{(Fig. \ref{fig:FluxHK}}, a \& b). { We find \hbox{$t_d = 0.0090 \pm 0.0038$} ($0.90 \pm 0.38$ per cent; significance 2.4$\sigma$) for \cahk and $t_d = 0.0089 \pm 0.0019$ ($0.89 \pm 0.19$ per cent; significance 4.8$\sigma$) for \ha.}
The transit depths for \transiti are highly sensitive to the few OOT observations, which show considerable scatter. A possible cause may be small flaring events. Observations of Proxima Centauri (M6V) have shown that following a flare event, elevated \cahk fluxes decay more rapidly than \ha \citep{fuhrmeister11proxcen}. If the outlying point in the \cahk light curve at $\phi = 0.0164$~(between $\phi_3$~and~$\phi_4$) is due to a small flare, it is possible that phases \hbox{($\phi =  0.0164,$}~\hbox{$0.0197$~\&~$0.02304$)} are all affected in the \ha light curve for \transiti. Obviously, more data points would have be{ en} desirable to verify this with greater statistical confidence. Nevertheless, removing the two OOT points at \hbox{$\phi = 0.0197$~\&~$0.02304$} from the \ha time series, which are used to normalise the light curve, we find \hbox{${ t_d = 0.0078 \pm 0.0021}$} (3.8$\sigma$). 
{ In \S \ref{section:results_mc} we perform a more robust empirical Monte Carlo simulation to better assess the significance of \transiti and \transitii estimates of $t_d$ \hbox{(Fig. \ref{fig:mc_all})}.}

\subsection{\transitiii light curves}
\protect\label{section:results_night3}

For \transitiii, the light curves do not permit accurate measurement of any excess absorption in \cahk and \ha owing to the rise in emission in the cores of both \cahk and \ha at the start of the transit. The elevated flux lasts for most of the transit, but appears to return to a level below that of the mean OOT phases somewhere in the $\phi_3\,-\,\phi_4$ interval. The transit depths shown in the bottom panels of Fig. \ref{fig:FluxHK} have been measured by taking the mean of two flux measurements, namely the minimum light curve flux in the $\phi_1\,-\,\phi_2$ region and the $\phi_3\,-\,\phi_4$ region. These measurements can be taken as lower limits to $t_d$ for \transitiii. However it is possible that variable flaring activity is present throughout the transit, which may result in an estimate of $t_d$ that is significantly underestimated.

\subsection{Significance of $t_d$ from empirical Monte Carlo}
\protect\label{section:results_mc}

To { better} assess the significance of the excess absorption during transit, we have performed an empirical Monte Carlo simulation in a similar manner to that initiated by \cite{redfield08}, and later adopted by \cite{astudillo-defru13}, \cite{cauley15hd189733} and \cite{wyttenbach15}. The key benefit of the simulations is that they enable the inclusion of atmospheric, instrumental and astrophysical uncertainties that are not fully characterised by the photon noise. The procedure is more critical in establishing the significance of \transiti than \transitii, where the excess absorption at transit phases is more clearly defined. Nevertheless, this procedure enables the propagated formal uncertainties that we have used so far (see Table \ref{table2}) to be assessed robustly. We did not attempt to assess the \transitiii data in this way.

Three scenarios are simulated, using (1) only OOT observations, (2) only IT observations and (3) IT and OOT observations. For the purposes of the simulation, we assumed IT observations were those observed in the range $\phi_2 \leq \phi \leq \phi_3$ from which we measured $t_d$ above. Similarly, the OOT observations are defined as phases where $\phi < \phi_1$ or $\phi > \phi_4$. For scenarios (1) and (2), half the observations were selected at random (without replacement) and then subdivided into two data sets in the same ratio as the actual observed IT and OOT observations. The transit depth, $t_d$, was calculated from the mean of the two simulated data subsets. The process was repeated 10,000 times, picking a random number of observations each time. The same procedure was adopted for scenario (3), but the two data subsets were drawn from the IT observations and the OOT observations separately. We expect that scenarios (1) and (2) should produce distributions that are consistent with \hbox{$t_d$ = 0}, while scenario (3) should yield the same, or similar values of $t_d$ to the measurements made on the full data sets.  

Fig. \ref{fig:mc_all} shows the resulting distributions of $t_d$ for the combined \cahk and \ha light curves for \transiti and \transitii. \cite{redfield08} and \cite{wyttenbach15} used the distribution width from scenario (1) using OOT observations only, as these phases are free of any transit effects. Since fewer data points are used compared with the measurement of $t_d$ from the full data set, the width of the scenario (2) distributions must be corrected by dividing the determined standard deviations of the simulated $t_d$ values by the square root of the ratio of the total IT+OOT observations to the OOT observations. For \cahk, we find respective \transiti and \transitii depths of $t_d = 0.0074 \pm 0.0044$ and $t_d = 0.0214 \pm 0.0022$ (i.e. 1.7$\sigma$ and 9.8$\sigma$ detections). { For \ha, we similarly find ${ t_d = 0.0084 \pm 0.0016}$ and ${ t_d = 0.0121 \pm 0.0012}$ { (i.e. 5.2$\sigma$ and 9.9$\sigma$ detections)}}. The empirical Monte Carlo transit depths and significance values are also shown in Fig. \ref{fig:mc_all} and are in good agreement with the measurements of $t_d$ in Table \ref{table2} (columns 4,5 \& 7) that use the propagated formal errors. The significance of the detections in each case are slightly reduced owing to excess variability above the photon noise level.

\begin{figure}
  \begin{center}
    \begin{tabular}{c}
      \hspace{-3mm} 
      \includegraphics[scale=0.365,angle=270]{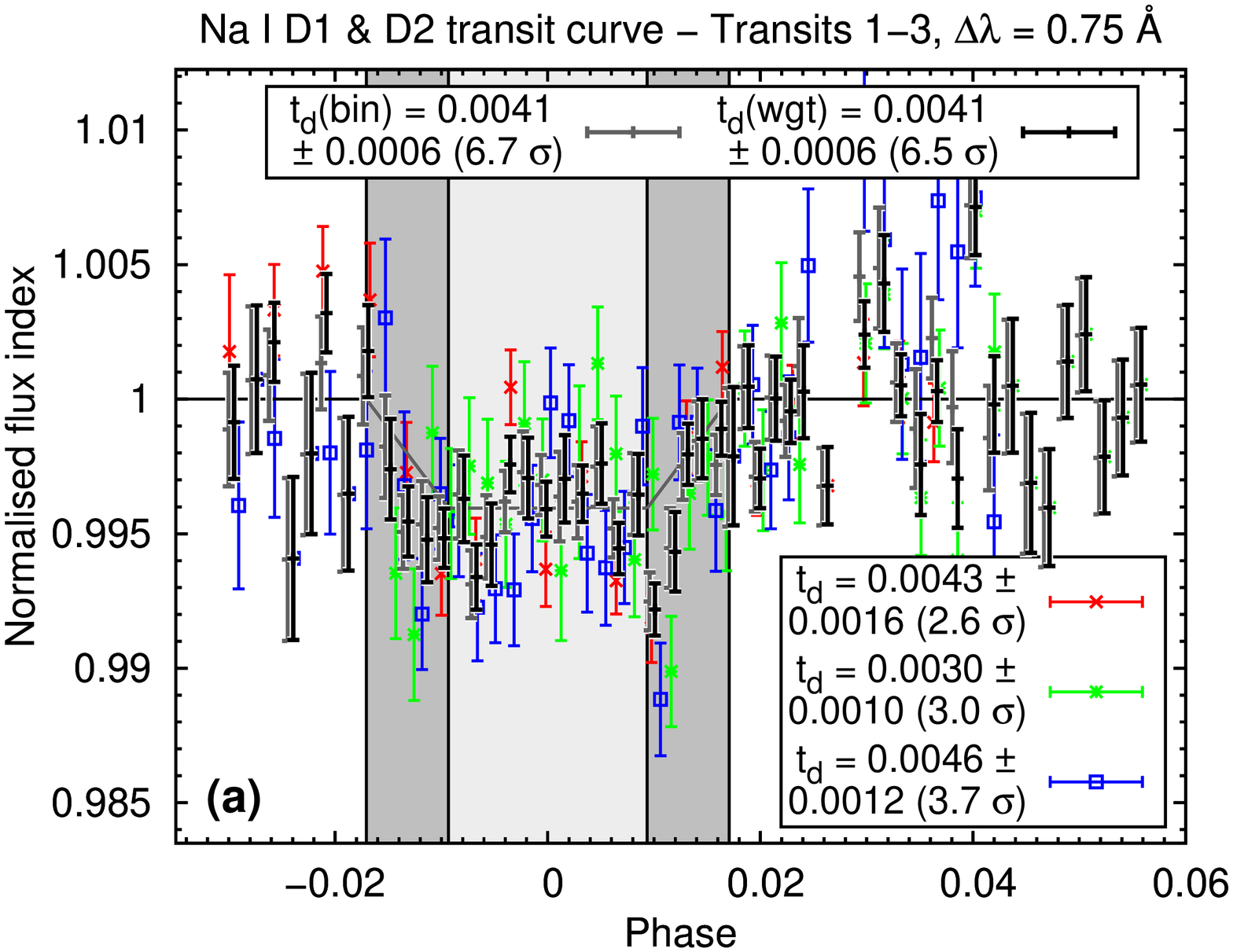} \\
      \hspace{-3mm}
      \includegraphics[scale=0.365,angle=270]{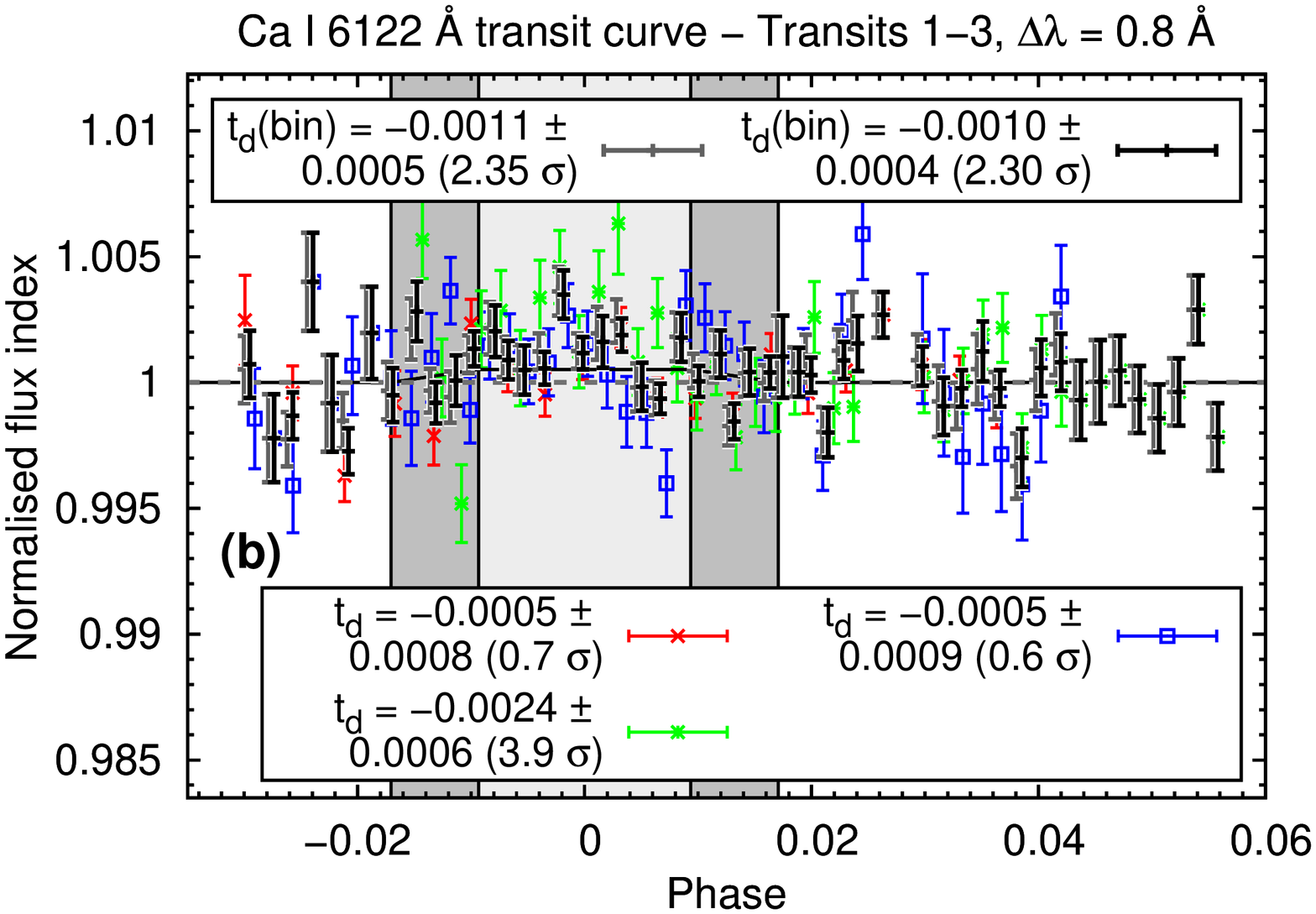} \\
    \end{tabular}
  \end{center}
  \caption{(a) Transit light curves for the combined Na 1 D1 (5995.924 \AA) and D2 (5889.951 \AA) lines. The mean light curves are shown in grey (unbinned) and black ($\Delta\phi=0.002$ bins). (b) Transit light curve for the \ionjb{Ca}{1} 6122 \AA\ line, plotted with the same axes as panel (a).\label{fig:FluxNa}}
\end{figure}

\subsection{\ionjb{Na}{1} D1 \& D2 and \ionjb{Ca}{1} 6122 \AA}
\protect\label{section:na1_ca1}

\subsubsection{\ionjb{Na}{1} D1 \& D2 transit light curves}
\protect\label{section:lightcurves_na1}

Since the depth of the excess absorption during transit appears to be variable from one epoch to another, we re-analysed the \nad lines at \hbox{5889.95 \AA}\ and \hbox{5895.92 \AA} for Transits 1\,-\,3 individually. As these lines have previously been studied in detail by other authors, and particularly, extensively and thoroughly by \cite{wyttenbach15} using the same data set as we are using, we adopted similar parameters throughout this paper. Here, we repeat the analysis, but also present the light curves for individual transits, as above for \cahk\ and \ha. For consistency with \cite{wyttenbach15}, we used 0.75 \AA\ passbands to derive the light curves, and have optimally combined the flux in both the D1 and D2 line cores in the same manner as for \cahk. 

Fig. \ref{fig:FluxNa} (a) shows the light curve for each transit and combined light curves using the data from all three transits, binned into $\Delta\phi = 0.001$ intervals. The {un-weighted}, binned, light curve is shown in grey, while the {weighted}, binned, light curve is shown in black. The weighted light curve using the three transits, is more appropriately adopted since it takes into account the measurement uncertainties in the individual transits. { We find ${ t_d = 0.0041 \pm 0.0006\ (6.5 \sigma)}$} for the weighted mean light curve for \nad. 

\begin{figure}
  \begin{center}
    \begin{tabular}{l}
      \hspace{-3mm}
      \includegraphics[scale=0.335,angle=270]{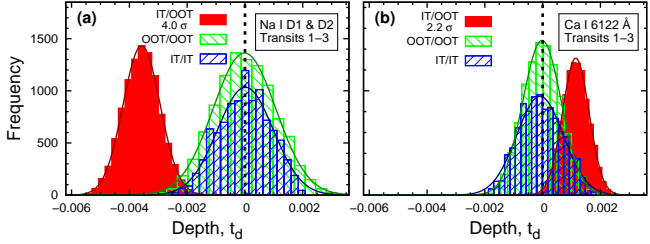} \\
    \end{tabular}
  \end{center}
  \caption{Empirical Monte Carlo analysis of the transit light curves for \nad and \ionjb{Ca}{1} 6122 \AA\ lines. \label{fig:mc_na1ca1}}
\end{figure}

{ Unlike \cahk and \ha, \transitii shows the shallowest depth in \nad. This result is not significant since all three transits agree within the uncertainties (see Fig. \ref{fig:FluxNa}). We note however that \cite{wyttenbach15} report an even shallower \transitii. Our measured $t_d$ using all three transits is marginally larger than the value reported by \cite{wyttenbach15} who find $t_d = 0.00325 \pm 0.00033$ when averaging Transits 1\,-\,3. Both results however agree within the $1\sigma$ measurement uncertainties. Fig. \ref{fig:mc_na1ca1} (a) shows the empirical Monte Carlo analysis applied to the weighted and binned \nad light curve. We find $t_d = 0.0036 \pm 0.0009$ ($4.0 \sigma$), bringing our estimate into closer agreement with that of \cite{wyttenbach15}.}

\subsubsection{Control light curve using \ionjb{Ca}{1} 6122 \AA}
\protect\label{section:lightcurves_ca1}

\ionjb{Ca}{1} 6122 \AA, a strong absorption line that does not contain a chromospheric component, was used by \cite{redfield08} as a control line. \cite{cauley15hd189733} similarly used three other \ionjb{Ca}{1} transitions as control lines. No transmission signature was found in these lines, in accordance with the findings of \cite{lodders99alkali} that \ionjb{Ca}{1} is only present in very low abundance in planetary atmospheres. We plot the light curve during transit of \ionjb{Ca}{1} 6122 \AA\ in Fig. \ref{fig:FluxNa} (b), { which shows that an obvious transit signature with significant absorption, as seen in the other lines, is not discernible.

Low level emission in \ionjb{Ca}{1} 6122 \AA\ is instead seen with marginal $2.3\sigma$ significance, with $t_d = -0.0010 \pm 0.0004$ (the negative sign represents relative emission during transit). The empirical Monte Carlo simulation in Fig. \ref{fig:mc_na1ca1} (b) { similarly indicates} $t_d = -0.0011 \pm 0.0005$ ($2.2\sigma$). This closer agreement between formal uncertainties and Monte Carlo estimates is likely due to the line being situated close to the centre of the \'{e}chelle order and the fact that it is a weaker line with more flux in the core. The S/N ratios of the line cores of \ionjb{Ca}{1} 6122 \AA\ for Transits 1\,-\,3 are $109.0$, $73.0$ and $64.5$ respectively, while for \nad, we find $65.8\ \&\ 67.0$,  $43.5\ \&\ 45.4$ and  $39.0\ \&\ 40.1$ respectively (c.f. also the S/N ratios for \cahk\ and \ha\ in \hbox{Table \ref{table1}}). 

The individual transits of \ionjb{Ca}{1} 6122 \AA\ in Fig. \ref{fig:FluxNa} (b) show marginally greater emission during \transitii, in agreement with the lower significance \transitii measurement for \nad. This finding may indicate that both lines are showing the signature of a starspot or starspot group during \transitii. The relative emission during transit may result from a combination of factors (see \S \ref{section:discussion}). The \hbox{\ionjb{Ca}{1} 6122 \AA}\ light curve demonstrates that not all lines exhibit excess absorption during transit.}

\begin{figure*}
\begin{center}
\includegraphics[scale=0.65,angle=270]{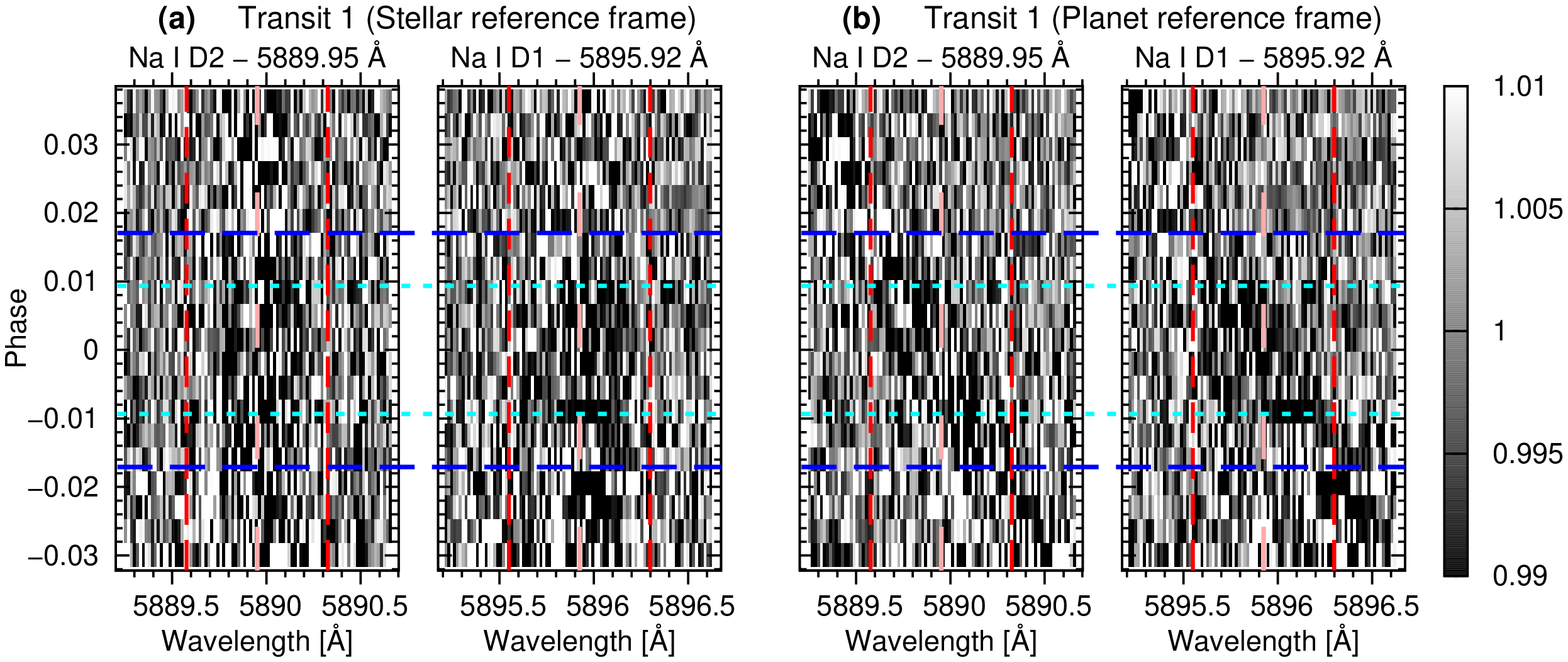} \\
\vspace{3mm}
\includegraphics[scale=0.65,angle=270]{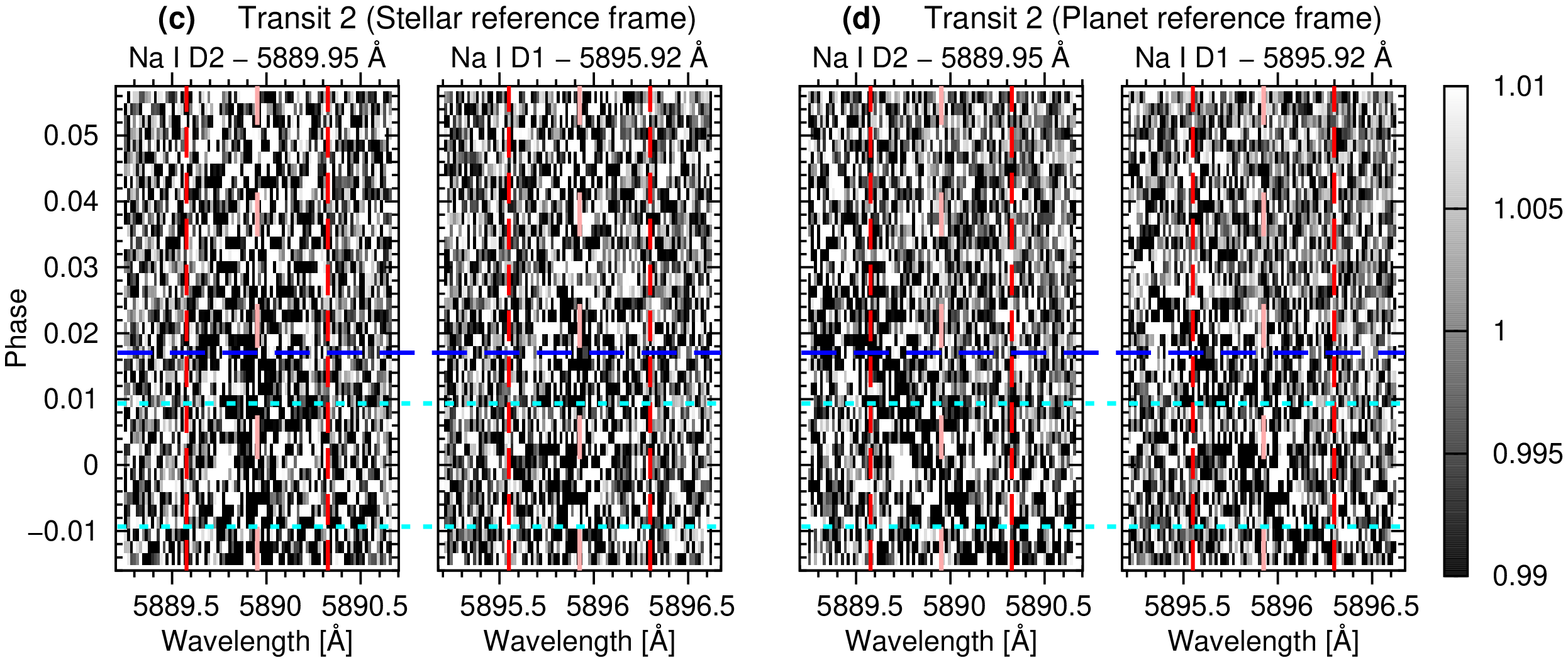} \\
\vspace{3mm}
\includegraphics[scale=0.65,angle=270]{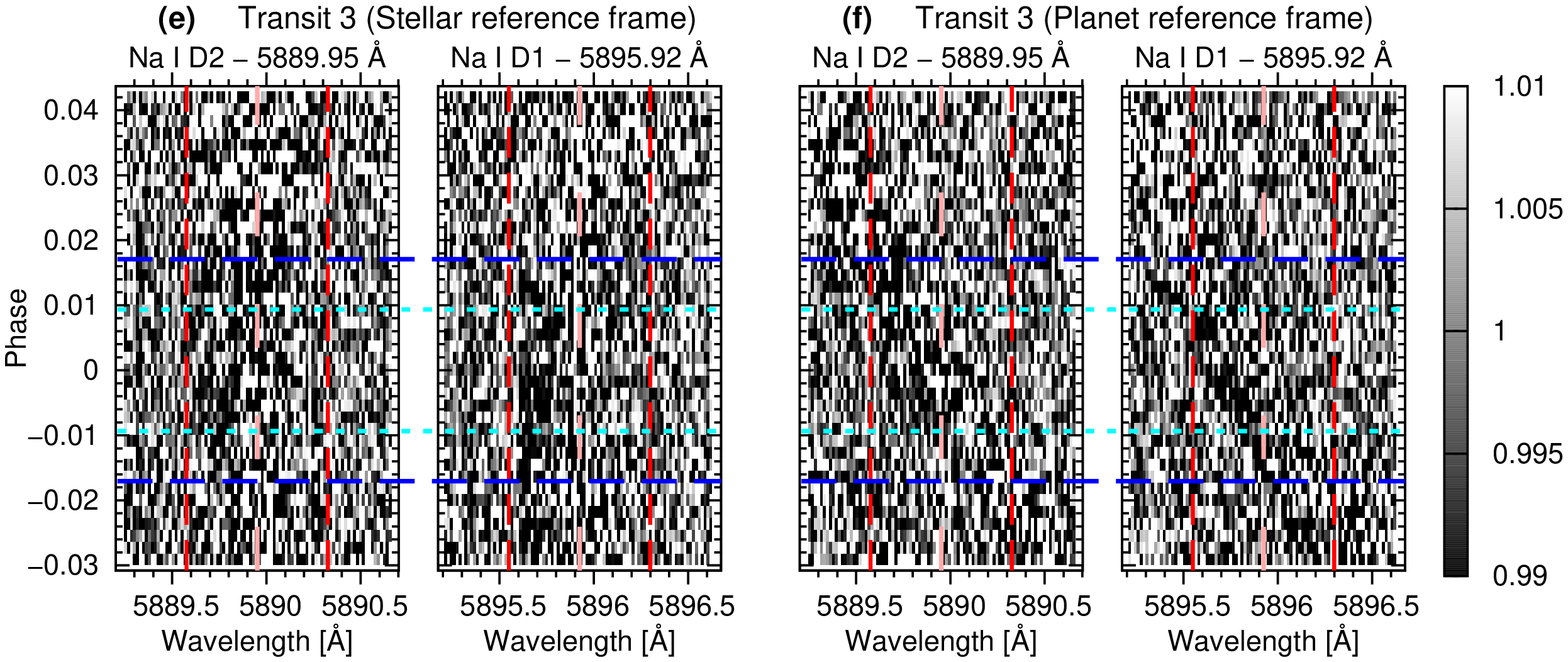} \\
\caption{As for Fig. \ref{fig:dynHK} with the \nad (5889.951 \AA and 3895.924 \AA) lines. The vertical dashed/red lines indicate the central $\pm 0.375$ \AA\ ($\Delta\lambda\ =\ 0.75$\ \AA) employed in other studies.\label{fig:dynNa}}
\end{center}
\end{figure*}

\subsection{Airmass effects}
\protect\label{section:airmass_effects}
{ We have considered the possibility that the observed excess absorption arises from incorrect normalisation or airmass effects at the time of observation. For the \transitii light curves presented in Fig. \ref{fig:FluxHK}, we have calculated Pearson's correlation coefficient to test for a correlation of the transit light curve fluxes with airmass, $X$, altitude above the horizon, $A$ (where $X = sec(A)$ for $A > 20$\degs), S/N ratio and seeing. We find respective correlation coefficients of $r = -0.63,\ 0.70,\ 0.55$ and $0.32$ for \cahk, while for \ha, we find $r = -0.75,\ 0.80,\ 0.48$ and $0.39$. The correlations with S/N ratio and seeing are moderate to weak, whereas moderate to high \hbox{(anti-)correlations} are found with airmass and altitude. To assess the significance of this finding, we calculated the correlation between a model transit signature and the variation in airmass and altitude at the observation phases on \transitii. We used the EXOFAST routine \citep{eastman13exofast} to create a model transit using the system parameters summarised in Table 1 of \citep{wyttenbach15}, but adopted U-band limb-darkening coefficients \citep{claret00ldc4} and the 3700\,-\,4200 \AA\ radius measured by \cite{sing11hd189733}. 
For the \cahk curve, the correlation between the simulated transit fit and airmass yields $r = -0.71$, while the correlation between the transit fit and altitude yields $r = 0.79$. 
The change in airmass and altitude with observation times thus show a similar strong correlation with an appropriate model transit light curve. 
A strong correlation between the observed light curves and airmass is thus not sufficient to rule out real transit effects. Moreover, given that \nad and \ionjb{Ca}{1} 6122 \AA\ do not show {significant} variability in depth during the three transits, we are confident that the time-variable light curves obtained for \cahk and \ha are the result of activity variability on \hd rather than airmass effects. Since there are also no telluric lines at the \cahk core wavelengths, we are not aware of any airmass effects that could give rise to the variable absorption signal seen in the line cores in \hbox{Fig. \ref{fig:IT_OOT}}.
}

\section{Time series and Transmission spectra}
\protect\label{section:results_transmission}

The residual time series spectra in Figs. \ref{fig:dynHK}, \ref{fig:dynHa} \& \ref{fig:dynNa} (a, c \& e) are plotted in the stellar reference frame. Below, we investigate the time series residual spectra in further detail and use them to obtain averaged residual transit profiles.

\subsection{Time series spectra in the planet reference frame}
\protect\label{section:timeseries_planet}

During transit, any stellar light absorbed in the atmosphere of the planet will be shifted by the instantaneous radial velocity of the planet relative to the star at the time of observation. In other words, the planetary absorption during transit should first appear at -15.9 \kms\ and at the end of transit will be shifted by +15.9 \kms. 

We have also included a correction for the stellar rotation of \hd. With \vsini = $3.1 \pm 0.03$ \kms \citep{colliercameron2010hd189733}, the impact parameter of the transit \citep{agol10hd189733} means that transmitted light is additionally shifted in the range $\pm 1.6$ \kms during the transit. The picture is also modified to a smaller degree by the Rossiter-MacLaughlin effect and putative zonal winds, as inferred by \cite{louden15hd189733} using the \transitiii\ observations of this date set. We have not included Rossiter-MacLaughlin velocity corrections here as they are negligible at $\sim$$30$ \ms \citep{winn06} compared with the other corrections. We follow the procedure adopted by \cite{wyttenbach15}, with the additional stellar rotation correction, to optimise the signal from the planet and avoid smearing due to its phase-dependent radial velocity. The instantaneous radial velocities were calculated from the ephemeris in \cite{agol10hd189733} using an orbital inclination of $i = 85.7$\degs.  

\subsubsection{\cahk and \ha planet reference frame residual time series}
\protect\label{section:timeseries_planet_ca_ha}

Figs. \ref{fig:dynHK} \& \ref{fig:dynHa} (panels b, d \& f) show the residual time series spectra in the planetary reference frame. For \cahk, as with panels a, c \& e (in the stellar reference frame), only the \transitii observations enable the absorption to be discerned by eye. For \ha the excess absorption is more readily { apparent} during Transits 1 \& 2 in Fig. \ref{fig:dynHa}, with a hint of some redshifted excess absorption during \transitiii (both panels e \& f) between $\phi_1$ and $\phi_4$. 

The most striking and unexpected feature of our analysis in this paper is the apparent red-to-blue trend of the excess absorption with phase. 
{ This appears when we shift to the reference frame moving with the planet. We expect absorption in the planet's atmosphere to be at rest in the planet's reference frame.
This is particularly clear during \transitii in the \ionjb{Ca}{2} H line and appears as a dark diagonal stripe during the transit. This is unexpected under} the assumption that the excess absorption arises in the atmosphere of the planet, we would expect improved alignment of the absorption in the planet reference frame. It is clear however that panel c of Fig. \ref{fig:dynHK}, appears to show no such gradient in the stellar reference frame. A red-to-blue trend with advancing phase is also evident in the planet reference frame time series panels for \ha in Fig \ref{fig:dynHa} (b, d, and possibly f despite the flaring event). Clearly this is a very important observation since it suggests that the excess absorption might arise in the stellar reference frame rather than the planetary reference frame.

\subsubsection{\nad residual time series}
\protect\label{section:timeseries_planet_na}

We plot time series spectra for the \nad lines for each transit in Fig. \ref{fig:dynNa} in both the stellar (panels a, c \& e) and planet (panels b, d \& f) reference frames. As with \ha, we have corrected for the effects of telluric lines. The stellar reference frame time series clearly show excess absorption during Transits 1\,-\,3. Since the significance of the excess absorption is smaller than for \cahk and \ha (\S \ref{section:results_night2}), the much lower S/N ratio in the line core regions is more evident, at both IT and OOT phases. The \nad lines are clearly also less sensitive to activity variability in the chromosphere as the \transitiii light curve still shows excess IT absorption unlike \cahk and \ha. In the planet reference frame (Fig. \ref{fig:dynNa}, panels b, d \& f), we see that the excess absorption shows a red-to-blue shift with phase for all three transits, again suggesting that the excess absorption may be stellar in origin. 

The time series spectra in both the stellar and planet reference frames (Fig. \ref{fig:dynNa}, panels a, c \& e) appear to show evidence that the excess absorption is sometimes redshifted (i.e. during \transiti) and sometimes blueshifted (\transitiii) from the rest wavelength. { For both \ha and \nad, the telluric line contributions are removed from the time series spectra to levels consistent with the noise before calculating the residual timeseries, as outlined in \S \ref{section:analysis_transit_ha}. A visual inspection of the telluric line strengths and positions, which shift against the stellar lines from one epoch to another, indicates that they are not responsible for the shift in position of the excess absorption.}

\subsection{Mean residual transit profiles}
\protect\label{section:transmission_spectrum}

Average residual IT spectra were created for observations  between contact phases $\phi_1$ and $\phi_4$. The first ground-based report of excess absorption in \nad by \cite{redfield08} obtained spectra by subtracting a master OOT spectrum from the IT spectra individually before co-adding the IT residuals to obtain a mean IT residual spectrum with optimal S/N ratio. Any resulting excess absorption at the wavelengths of the atomic species being studied has been interpreted as a signature of the planet, without assessing the time-dependent velocity behaviour of the absorption. In contrast, and in light of our analysis in \S \ref{section:timeseries_planet},  we have created transmission profiles for \cahk, \nad and \ha in both the stellar reference frame and the planet reference frame.

Fig. 9 shows transmission profiles for each line for Transits 1\,-\,3 in both the stellar and planet reference frames, as indicated. The plots maintain the same scales for a given line in both reference frames, and for each transit. A Gaussian profile has been fitted to each spectrum to obtain an estimate of the velocity offset of the feature, $v$ (\kms), relative to the rest wavelength of each line, the Gaussian width, $w$ (in \kms), and the normalised depth, $d$. The three parameters, along with formal uncertainties, are indicated in each panel. A positive value of $d$ indicates absorption. Since \transitiii\ is affected by excess chromospheric activity, and we do not detect excess absorption in \cahk, it is not surprising that meaningless fits are obtained. As expected from our findings in \S \ref{section:results_lightcurves}, all lines show the strongest absorption during \transitii.

\subsubsection{Transmission profile differences in the stellar and planet reference frame}
\protect\label{section:transmission_spectrum_significance}

Fig. \ref{fig:trans_sig} (a) displays the measured $v$, $d$ and $w$, for each line in the stellar (open symbols) and planet (closed symbols) reference frames for each line. In Fig. \ref{fig:trans_sig} (b), the relative difference significance of the planet minus the stellar reference frame measurement (e.g. for the profile width, $w$, $\sigma(w_p-w_s)$ = $(w_p - w_s) / \sqrt{\Delta w^2_p - \Delta w^2_s}$) is plotted for each parameter. { \transiti (red sqaures) is only marginally detected in most lines. Consistent with this, the difference between the transit line profile in the two frames of reference is not significant: for \transiti, only \ionjb{Na}{1} D2 shows $\sigma(w_p-w_s)$ and $\sigma(d_p-d_s)$\ $> 1$, while \ionjb{Na}{1} D1 shows respective significances of $\sim 1$.}
For the \cahk and \ha lines, any differences are very low confidence, with  $|\sigma(w_p-w_s)| < 1$ and $|\sigma(d_p-d_s)| < 1$. In other words, marginally narrower and deeper profiles are seen {\em in the stellar reference frame} during \transiti only for the less chromospherically sensitive \nad lines. 

{ In Fig. 9, the \transitii plots (columns 3 and 4) reveal that {\em all profiles are deeper and narrower in the stellar reference frame}}, although the significance is { still} lower in \ionjb{Ca}{2} K and \ha. For \ionjb{Ca}{2} H, \nad, $1 \lesssim \sigma(w_p-w_s) \lesssim 3$. For \transitiii, \nad\ show similar results to \transiti on average. While the \cahk lines are strongly affected by the chromospheric activity transient, the \ha\ absorption profile remains marginally sharper in the stellar reference frame, but with low significance, as at all other epochs, { likely due to the fact that the feature is broader than the other lines. Finally, the significance of the difference between the reference frames of the velocity offsets from the rest wavelengths, $\sigma(v_p-v_s)$, is potentially less important, but we include it for completeness}.

\begin{figure*}
\vbox to220mm{
\includegraphics[scale=0.995,angle=0]{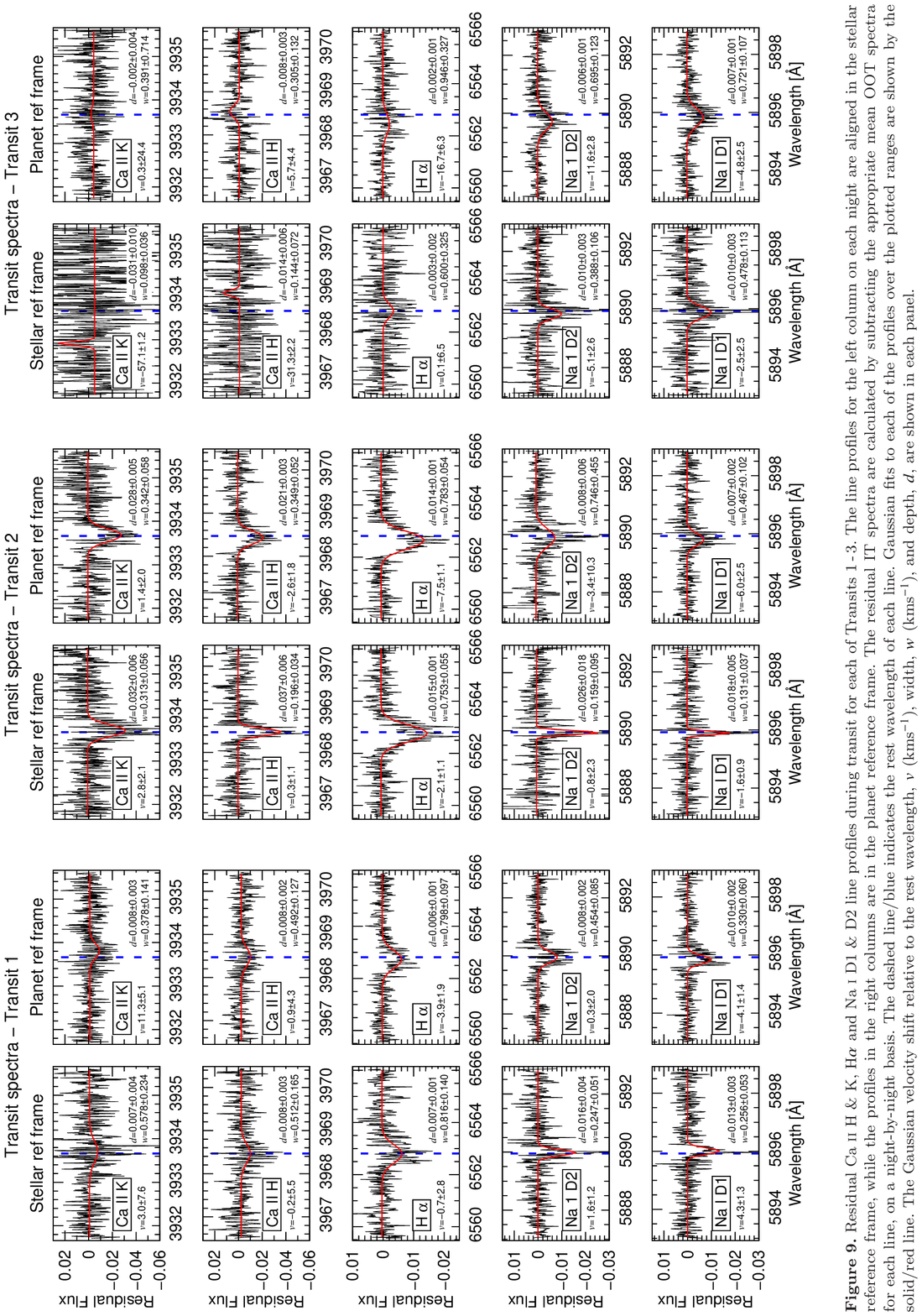}
\vfil}
\protect\label{fig:transmission_spectra_all}
\end{figure*}

\setcounter{figure}{9}

\begin{figure*}
  \begin{center}
    \begin{tabular}{l}
      \hspace{-3mm}
      \includegraphics[scale=0.55,angle=270]{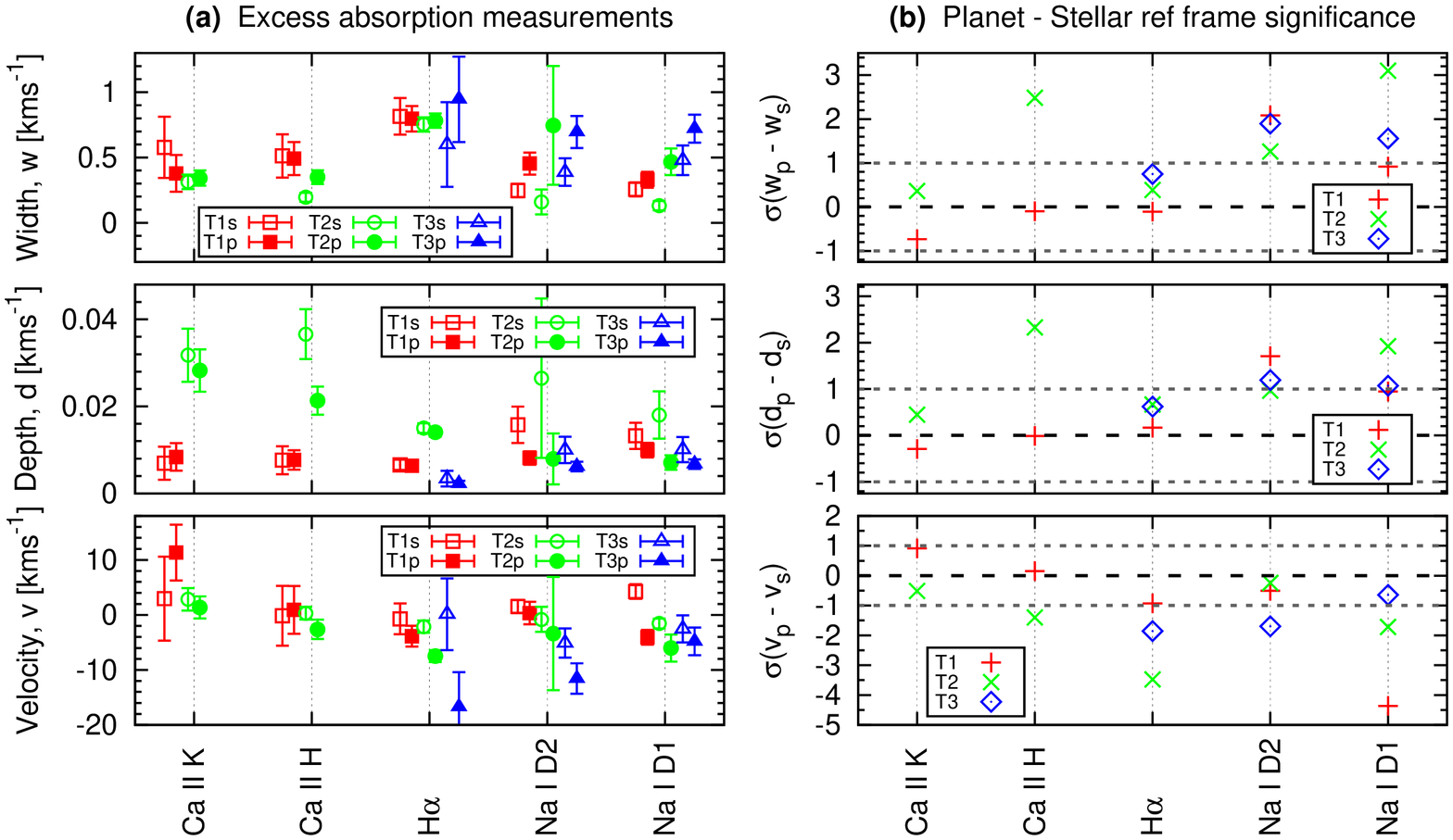} \\
    \end{tabular}
  \end{center}
  \caption{(a) The transit spectra depth, $d$, width, $w$ (\kms), and velocity shift, $v$ (\kms), are shown for each line, \cahk, \ha\ and \nad, and for each Transit 1\,-\,3 in the stellar (labelled T1s\,-\,T3s, open symbols) and planet (labelled T1p\,-\,T3p, closed symbols) reference frames. The \cahk\ points for \transitiii\ are not plotted. Right panels: The difference between the measurement of each parameter in the planet reference frame minus the stellar frame as the significance determined from the uncertainties in the measurements. The zero significance (long dashed) and $\pm 1\sigma$ (short dashed) levels are plotted. \protect\label{fig:trans_sig}}
\end{figure*}

\section{Discussion}
\protect\label{section:discussion}

Excess absorption is seen in some or all of the \cahk, \ha and \nad lines during all three transits. The sensitivity of the \cahk lines to chromospheric transients are a notable obstacle to detecting transit signatures in these lines, as evidenced by the \transitiii observations. This problem was encountered by \cite{czesla15hd189733} in their study of the centre-to-limb brightening of \cahk and \nad with high S/N ratio UVES observations.

Despite extending and confirming previous results, our study reveals some surprises. Firstly, from space-based observations, it is not clear that \ionjb{Ca}{2} should show strong excess absorption in the atmosphere of a close orbiting giant exoplanet such as \hdone. Observations by \cite{pont08hd189733} and \cite{sing11hd189733} with the Hubble Space Telescope (HST) did not initially detect either Ca or Na.
\cite{pont13hd189733} were able to identify Na absorption by adopting a narrower passband, but at a lower level than predicted. The detection of relatively weak absorption only in narrow regions centred on the Na lines added weight to the earlier suggestion \citep{etangs08rayleigh} that a Rayleigh scattering haze  is responsible for obscuring all but strong, sharp, line cores. While \cite{sing11hd189733} and \cite{pont13hd189733} demonstrate that a haze is detected at short wavelengths, masking planetary spectral features, the cloud-free model transmission spectra of \cite{fortney10transmission} {\em do} lead us to expect absorption from the \cahk\ lines (e.g. see also Figs. 10 \& 11 in \citealt{pont13hd189733}). With the increasing importance of Rayleigh scattering at shorter wavelengths and the lower S/N ratios, it is not surprising that detection at \cahk was not made by \cite{sing11hd189733} or \cite{pont13hd189733}.

The ground-based observations of \cite{redfield08} and \cite{wyttenbach15} have used high resolution spectroscopy and similarly show excess absorption in the cores of Na during transit of \hdone. Other studies have also claimed excess absorption in the Hydrogen Balmer lines of { \hdone} \citep{cauley15hd189733}. In this paper, we also find variable absorption in broad agreement with the \nad results in \cite{wyttenbach15}, but further, we find that absorption strength is variable from transit to transit in all the lines studied, { though the significance in \nad is uncertain.} This suggests either that the structure of the upper layers of the planetary atmosphere may be variable, or that stellar variability to some degree may instead be responsible for the excess absorption.

In addition to the variable nature of the transits, we find that the excess absorption is narrowest and deepest when the time resolved spectra are co-added in the stellar reference frame. { This is clear and obvious in the ten panels of Fig. 9 showing results from \transitii.} Although the significance { of the differences between excess absorption profile widths and depths, $\sigma(w_p-w_s)$ and $\sigma(d_p-d_s)$ (\S \ref{section:transmission_spectrum_significance})}, are relatively low for some lines (i.e. \ionjb{Ca}{2} K and \ha), the \nad lines taken together consistently appear sharper in the stellar reference frame. This is probably due to the narrow \nad line cores. { Similarly, the \ionjb{Ca}{2} H profile appears significantly sharper in the stellar reference than the planetary reference frame (as quantified by $\sigma(w_p-w_s)$ and $\sigma(d_p-d_s)$) during \transitii, whereas \ionjb{Ca}{2} K does not because it is intrinsically broader as explained in \S \ref{section:results_night2}.}

For the sharpest features in Figs. \ref{fig:dynHK}, \ref{fig:dynHa} \& \ref{fig:dynNa}, we have further verified that the apparent red-to-blue shift of the excess absorption in the planet reference frame is real. We summed residual spectra in the planet reference frame between $\phi_1$ and $\phi_2$ (Profile 1) and also between $\phi_3$ and $\phi_4$ (Profile 2). Within the measurement uncertainties, the difference between the measured velocities of Profile 1 and Profile 2 is identical to the phase dependent shift applied to translate the spectra into the planet reference frame. This further confirms our suspicion that the excess absorption, or at least the more significant part of it, is located in the stellar reference frame.

\subsection{The reliability of transmission spectra in chromospherically sensitive lines}

\protect\label{section:discussion_reliability}

Ground-based transmission spectroscopy using high-resolution spectrometers must normalise the flux to account for the non-constant flux levels from seeing and instrumental effects. On the face of it, the simple procedure of using continuum regions on either side of the line seems reasonable as it is able to provide the precision needed to detect the apparent excess absorption with significance. This is a doubly differential method of transmission spectroscopy where first the line is compared with nearby continuum regions and then the IT and OOT spectra are differenced. We postulate this method can lead to either spurious transit signatures, or at least systematically biased transit signatures.
{ We might similarly expect that space-based measurements, although free from ground-based atmospheric effects, would be susceptible to chromospheric variability. However passbands of typically a few hundred \AA\ that include continuum and photospheric lines have typically been employed in space-based observations, thereby diluting the chromospheric contribution, which is confined to narrow line cores.

The problem with normalisation lies with the assumption that the flux in the continuum and the flux in the line cores arise from the same region of the star. This is obviously not the case in the strong lines we are interested in, which all contain a chromospheric emission component.
The adopted continuum regions used for normalisation arise in the stellar photosphere, although they also include weaker photospheric lines, especially in the densely populated wavelength regions of \cahk.
The transit index measurement (Equations \ref{equation:HK_flux} \& \ref{equation:halpha_flux}) is therefore only strictly applicable to lines that arise from a similar part of the atmosphere to the continuum regions, and even then only for an immaculate, unspotted photosphere. 
If \hdone\ possesses active regions confined to specific areas of the star, the transit index measurement may appear to show excess absorption during the transit, since blocking any active regions that contribute to the line core emission creates a larger contrast effect than blocking the photosphere (c.f. Chapter 5 of \citealt{haswell2010book}).

In chromospherically sensitive lines such as \hdone, where log $R'_{HK}$ exceeds solar maximum by 0.35 dex in log $R'_{HK}$ (a factor of 2.2), active regions are thus likely to interfere with calculation of the light curves. Starspot corrections are generally required for broadband transmission spectrophotometry \citep{pont08hd189733,sing11hd189733}, but starspot effects may be more localised to instances where the planet transit crosses spots. Active regions, producing chromospheric emission may effect transit light curves in a similar way, but the chromospheric component in line cores probably systematically affects the whole transit lightcure depth.
This can be seen clearly in the \hd \cahk\ and \ha light curves that show a deeper \transitii compared with \transiti. 
he cores in each line studied here are also sensitive to different regions of the chromosphere and show different sensitivities to flares and other transients.
Our analysis of the \nad lines shows variability, but in fact \transitii is {\em less deep} than Transits 1 \& 3, although the results agree within the uncertainties. \cite{wyttenbach15} also find variable transit depths in \nad, measuring $0.00339 \pm 0.00057$ for \transiti, $0.00143 \pm 0.00059$ for \transitii and $0.00457 \pm 0.00066$ for \transitiii. { Our results for \nad are thus consistent with \cite{wyttenbach15} except for the \transitii measurement.}
The evidence for variability, {\em due specifically to chromospheric variability} in the \nad cores, is thus less clear than for \cahk and \ha, and may be the result of lower sensitivity of the \nad lines to chromospheric activity.

One explanation for the very slight excess emission we find during IT phases for the photospheric \ionjb{Ca}{1} \hbox{6122 \AA}\ line might be the presence of cool starspots (e.g. see \citealt{pont08hd189733}) associated with active regions on \hd. The eclipsing of starspots could explain the greater degree of emission seen during \transitii. \cite{sing11hd189733} find corrections for starspots, that are variable from transit to transit, of order 1 per cent (an order of magnitude greater than the precision of their transit depth measurements) are necessary for \hdone. { The shallower transit we see in \nad may therefore also be partially due to the presence of cool starspots. Variability in the Na lines could thus arise from both photospheric starspots and chromospheric plage regions.}

It is also possible that we have detected the centre-to-limb variation (CLV) due to limb-darkening of the stellar disk and changes in the line profile shapes as a function of limb angle. Both the \nad and \ionjb{Ca}{1} light curves in Fig. \ref{fig:FluxNa} show marginal evidence for a bump in the transit region, with a morphology similar to that modelled and found observationally by \cite{czesla15hd189733}. For \hdone transits, \cite{czesla15hd189733} found CLV in regions {\em around} the \nad cores at the few $\times\,10^{-3}$ level (unfortunately the core regions were excluded owing to a flaring event during those observations). In addition, the CLV of the earth in strong solar lines was investigated by \cite{yan15clv} by taking spectra of the eclipsed moon. They demonstrated that for the Earth-Sun transits, the effect is more pronounced if narrow passbands are adopted when calculating the transit curves. This is important because studies that assert that the excess absorption signal is planetary in origin, find the strongest absorption in the narrow core regions of the \nad lines. The magnitude of the CLV, specifically in the core regions of strong absorption lines, clearly requires further investigation in stars such as \hd with close orbiting gas giant planets.}

\subsection{\cahk vs \ha chromospheric signatures}
\protect\label{section:discussion_activity}

The observation of deep transit signatures from \cahk may be explained by contrast effects, since { chromospheric} emission in the \cahk { cores} is stronger as a fraction of nearby continuum flux than in \ha and \nad. \cite{livingston07chrom} showed that a correlation exists between { chromospheric emission in} \cahk\ and other lines, including \ha, during the solar cycle. The amplitude of variability in \cahk\ was found to be stronger than in \ha. This was also shown in Fig. 1 of \cite{meunier09chrom}, which demonstrates that \cahk\ core emission is an order of magnitude more variable than \ha\ across a solar magnetic cycle. \cahk and \ha are also affected by different phenomena and do not arise from the same region of the chromosphere (e.g. see \citealt{vernazza81} and review by \citealt{hall08chrom}). While both \cahk and \ha arise from plage regions, \ha is also more sensitive to absorption from filaments, potentially reducing the observed line core emission. With greater core emission in \cahk\ and a higher degree of variability between quieter and more active periods on the Sun, we would thus expect to see both deeper apparent transits and more variability in \cahk than in \ha.  
{ The extent to which the solar observations can be extrapolated} to a more active K star such as \hd is uncertain. Significant scatter was found in similar correlations in a sample of FGK and M stars, following earlier suggestions that such correlations exist (\citealt{cincunegui07activity} and references therein).

\subsection{Excess absorption velocities}
\protect\label{section:discussion_velocities}

The excess absorption seen during transit also shows variable velocity offsets. This has been interpreted as winds in the upper atmosphere of \hdone with estimates up to $\sim 8$ \kms \citep{wyttenbach15}. \cite{louden15hd189733} used the \transitiii data to spatially resolve the Na absorption, reporting winds on the leading and trailing limb of the planet, which they interpreted as an eastward equatorial jet. However, we detect a transient increase in line core flux in \cahk and \ha during \transitiii, probably the result of a small flare. This means it is likely that any putative planetary wind measurements will be severely affected by chromospheric/coronal transients. Our time series spectra in Fig. \ref{fig:dynNa} (e)  indicate excess absorption aligned in the stellar reference frame during \transitiii, with velocities of $-5.1 \pm 2.6$ \kms (\ionjb{Na}{1} D1) and $-2.5 \pm 2.5$ \kms (\ionjb{Na}{1} D2) (Figs. \ref{fig:transmission_spectra_all} \& \ref{fig:trans_sig}). { We conclude the putative planetary wind signature in fact arises in the dynamic behaviour of the outer layers of the stellar atmosphere.}

\section{Summary \& Conclusion}
\protect\label{section:conclusion}

We have detected excess absorption in the \cahk lines of \hd during transit of its close orbiting giant exoplanet. We also detected excess absorption in \ha, and re-examined previous results for \nad. The transit depths are significantly variable { in \cahk and \ha, with \transitii exhibiting the strongest line absorption. We} studied the time series spectra of each line during the three transits and find evidence that the excess absorption is located in the stellar reference frame. We postulate that the chromosphere of \hd is the dominant contributor to the excess transit absorption and responsible for its variable nature 
{ since the flux measured in the cores of the lines we have studied contains a significant chromospheric contribution. A systematic affect is introduced because for any given transit, the chromospheric emission will be slightly different, and localised to specific regions on the stellar surface, whereas the continuum emission used for normalisation originates in the stellar photosphere. Temporal variability (from one transit to the next) in chromospheric emission will lead to changes in the depth of the apparent absorption, while localised systematic effects, where the planet transits active regions, will also result in variability. The \nad lines core are less sensitive to chromospheric variability than \cahk and \ha and may contain systematics from both chromospheric features such as plage and filaments, and photospheric variability from starspots.}

Observations aimed at detecting { specific lines in} the atmospheres of transiting exoplanets with ground-based observations thus likely yield systematically biased transit signatures { if the lines being probed are sensitive to stellar chromospheric emission}. Space-based observations can, in principle, measure the flux time series for all observed wavelengths so that (singly) differential transmission spectroscopy reveals the difference between IT and OOT spectra. In practice however, often the OOT { spectra} are used to remove systematic effects, or detrending is necessary \citep{pont07hd189733} so even space-based transmission spectroscopy may be affected by this systematic bias when observing chromospherically sensitive regions \citep{pont13hd189733}. It is worth noting also that { since active regions vary in time, the issue is likely to be} more important for planet hosting stars with higher activity levels, such as \hd.

Our analysis shows that multiple transits and careful scrutiny of time series spectra are desirable to assess the nature of excess absorption during exoplanetary transits. It would be interesting to obtain further transit observations of both \hd and \hbox{HD 209458} in combination with long term activity monitoring. Study of \cahk is both difficult and risky because it requires that the star be active, with sufficient line reversal flux to obtain reliable transit light curves. \hd is the most favourable target for study of \cahk transit effects; even \hbox{HD 209458}, being much less active, does not possess significant \cahk line core flux. The \nad and \ha lines occur in regions with better continuum flux and thus offer the potential to investigate transit and stellar activity effects on a wider sample of systems.

\section*{Acknowledgments}
J.R.B. and C.A.H. were supported by the STFC under the grant ST/L000776/1. D.S. was supported by an STFC studentship. We would like to thank the anonymous referee for reading and providing constructive feedback to help improve the manuscript.





\begin{thebibliography}{}
\makeatletter
\relax
\def\mn@urlcharsother{\let\do\@makeother \do\$\do\&\do\#\do\^\do\_\do\%\do\~}
\def\mn@doi{\begingroup\mn@urlcharsother \@ifnextchar [ {\mn@doi@}
  {\mn@doi@[]}}
\def\mn@doi@[#1]#2{\def\@tempa{#1}\ifx\@tempa\@empty \href
  {http://dx.doi.org/#2} {doi:#2}\else \href {http://dx.doi.org/#2} {#1}\fi
  \endgroup}
\def\mn@eprint#1#2{\mn@eprint@#1:#2::\@nil}
\def\mn@eprint@arXiv#1{\href {http://arxiv.org/abs/#1} {{\tt arXiv:#1}}}
\def\mn@eprint@dblp#1{\href {http://dblp.uni-trier.de/rec/bibtex/#1.xml}
  {dblp:#1}}
\def\mn@eprint@#1:#2:#3:#4\@nil{\def\@tempa {#1}\def\@tempb {#2}\def\@tempc
  {#3}\ifx \@tempc \@empty \let \@tempc \@tempb \let \@tempb \@tempa \fi \ifx
  \@tempb \@empty \def\@tempb {arXiv}\fi \@ifundefined
  {mn@eprint@\@tempb}{\@tempb:\@tempc}{\expandafter \expandafter \csname
  mn@eprint@\@tempb\endcsname \expandafter{\@tempc}}}

\bibitem[\protect\citeauthoryear{{Agol}, {Cowan}, {Knutson}, {Deming},
  {Steffen}, {Henry}  \& {Charbonneau}}{{Agol} et~al.}{2010}]{agol10hd189733}
{Agol} E.,  {Cowan} N.~B.,  {Knutson} H.~A.,  {Deming} D.,  {Steffen} J.~H.,
  {Henry} G.~W.,   {Charbonneau} D.,  2010, \mn@doi [ApJ]
  {10.1088/0004-637X/721/2/1861}, \href
  {http://cdsads.u-strasbg.fr/abs/2010ApJ...721.1861A} {721, 1861}

\bibitem[\protect\citeauthoryear{{Astudillo-Defru} \& {Rojo}}{{Astudillo-Defru}
  \& {Rojo}}{2013}]{astudillo-defru13}
{Astudillo-Defru} N.,  {Rojo} P.,  2013, \mn@doi [A\&A]
  {10.1051/0004-6361/201219018}, \href
  {http://cdsads.u-strasbg.fr/abs/2013A%26A...557A..56A} {557, A56}

\bibitem[\protect\citeauthoryear{{Barman}}{{Barman}}{2007}]{barman07}
{Barman} T.,  2007, \mn@doi [ApJ] {10.1086/518736}, \href
  {http://cdsads.u-strasbg.fr/abs/2007ApJ...661L.191B} {661, L191}

\bibitem[\protect\citeauthoryear{{Barnes} et~al.,}{{Barnes}
  et~al.}{2012}]{barnes12rops}
{Barnes} J.~R.,  et~al., 2012, \mn@doi [MNRAS]
  {10.1111/j.1365-2966.2012.21236.x}, \href
  {http://cdsads.u-strasbg.fr/abs/2012MNRAS.424..591B} {424, 591}

\bibitem[\protect\citeauthoryear{{Boisse} et~al.,}{{Boisse}
  et~al.}{2009}]{boisse08hd189733}
{Boisse} I.,  et~al., 2009, \mn@doi [A\&A] {10.1051/0004-6361:200810648}, \href
  {http://cdsads.u-strasbg.fr/abs/2009A%26A...495..959B} {495, 959}

\bibitem[\protect\citeauthoryear{{Bouchy} et~al.,}{{Bouchy}
  et~al.}{2005}]{bouchy05hd189733}
{Bouchy} F.,  et~al., 2005, A\&A, 444, L15

\bibitem[\protect\citeauthoryear{{Cauley}, {Redfield}, {Jensen}, {Barman},
  {Endl}  \& {Cochran}}{{Cauley} et~al.}{2015}]{cauley15hd189733}
{Cauley} P.~W.,  {Redfield} S.,  {Jensen} A.~G.,  {Barman} T.,  {Endl} M.,
  {Cochran} W.~D.,  2015, \mn@doi [ApJ] {10.1088/0004-637X/810/1/13}, \href
  {http://cdsads.u-strasbg.fr/abs/2015ApJ...810...13C} {810, 13}

\bibitem[\protect\citeauthoryear{{Charbonneau}, {Brown}, {Latham}  \&
  {Mayor}}{{Charbonneau} et~al.}{2000}]{charbonneau00transit}
{Charbonneau} D.,  {Brown} T.~M.,  {Latham} D.~W.,   {Mayor} M.,  2000, \mn@doi
  [ApJ] {10.1086/312457}, \href
  {http://adsabs.harvard.edu/cgi-bin/nph-bib_query?bibcode=2000ApJ...529L..45C&db_key=AST}
  {529, L45}

\bibitem[\protect\citeauthoryear{{Charbonneau}, {Brown}, {Noyes}  \&
  {Gilliland}}{{Charbonneau} et~al.}{2002}]{charbonneau02atmos}
{Charbonneau} D.,  {Brown} T.~M.,  {Noyes} R.~W.,   {Gilliland} R.~L.,  2002,
  \mn@doi [ApJ] {10.1086/338770}, \href
  {http://adsabs.harvard.edu/cgi-bin/nph-bib_query?bibcode=2002ApJ...568..377C&db_key=AST}
  {568, 377}

\bibitem[\protect\citeauthoryear{{Cincunegui}, {D{\'{\i}}az}  \&
  {Mauas}}{{Cincunegui} et~al.}{2007}]{cincunegui07activity}
{Cincunegui} C.,  {D{\'{\i}}az} R.~F.,   {Mauas} P.~J.~D.,  2007, \mn@doi
  [A\&A] {10.1051/0004-6361:20066503}, \href
  {http://cdsads.u-strasbg.fr/abs/2007A%26A...469..309C} {469, 309}

\bibitem[\protect\citeauthoryear{{Claret}}{{Claret}}{2000}]{claret00ldc4}
{Claret} A.,  2000, A\&A, 363, 1081

\bibitem[\protect\citeauthoryear{{Clough}, {Iacono}  \& {Moncet}}{{Clough}
  et~al.}{1992}]{clough92}
{Clough} S.~A.,  {Iacono} M.~J.,   {Moncet} J.~L.,  1992, {J. Geophys. Res.},
  97, 15761

\bibitem[\protect\citeauthoryear{{Clough}, W., {Mlawer}, {Delamere}, {Iacono},
  {Cady-Pereira}, {Boukabara}  \& {Brown}}{{Clough} et~al.}{2005}]{clough05}
{Clough} S.~A.,  W. S.~M.,  {Mlawer} E.~J.,  {Delamere} J.~S.,  {Iacono} M.~J.,
   {Cady-Pereira} K.,  {Boukabara} S.,   {Brown} P.,  2005, {J. Quant.
  Spectrosc. Radiat. Transfer}, 91, 233

\bibitem[\protect\citeauthoryear{{Collier Cameron}, {Horne}, {Penny}  \&
  {Leigh}}{{Collier Cameron} et~al.}{2002}]{cameron02upsand}
{Collier Cameron} A.,  {Horne} K.,  {Penny} A.,   {Leigh} C.,  2002, MNRAS,
  330, 187

\bibitem[\protect\citeauthoryear{{Collier Cameron}, {Bruce}, {Miller}, {Triaud}
   \& {Queloz}}{{Collier Cameron} et~al.}{2010}]{colliercameron2010hd189733}
{Collier Cameron} A.,  {Bruce} V.~A.,  {Miller} G.~R.~M.,  {Triaud}
  A.~H.~M.~J.,   {Queloz} D.,  2010, \mn@doi [MNRAS]
  {10.1111/j.1365-2966.2009.16131.x}, \href
  {http://cdsads.u-strasbg.fr/abs/2010MNRAS.403..151C} {403, 151}

\bibitem[\protect\citeauthoryear{{Czesla}, {Klocov{\'a}}, {Khalafinejad},
  {Wolter}  \& {Schmitt}}{{Czesla} et~al.}{2015}]{czesla15hd189733}
{Czesla} S.,  {Klocov{\'a}} T.,  {Khalafinejad} S.,  {Wolter} U.,   {Schmitt}
  J.~H.~M.~M.,  2015, \mn@doi [A\&A] {10.1051/0004-6361/201526386}, \href
  {http://cdsads.u-strasbg.fr/abs/2015A%26A...582A..51C} {582, A51}

\bibitem[\protect\citeauthoryear{{Eastman}, {Gaudi}  \& {Agol}}{{Eastman}
  et~al.}{2013}]{eastman13exofast}
{Eastman} J.,  {Gaudi} B.~S.,   {Agol} E.,  2013, \mn@doi [PASP]
  {10.1086/669497}, \href {http://cdsads.u-strasbg.fr/abs/2013PASP..125...83E}
  {125, 83}

\bibitem[\protect\citeauthoryear{{Fortney}, {Sudarsky}, {Hubeny}, {Cooper},
  {Hubbard}, {Burrows}  \& {Lunine}}{{Fortney}
  et~al.}{2003}]{fortney03hd209458}
{Fortney} J.~J.,  {Sudarsky} D.,  {Hubeny} I.,  {Cooper} C.~S.,  {Hubbard}
  W.~B.,  {Burrows} A.,   {Lunine} J.~I.,  2003, \mn@doi [ApJ]
  {10.1086/374387}, \href {http://cdsads.u-strasbg.fr/abs/2003ApJ...589..615F}
  {589, 615}

\bibitem[\protect\citeauthoryear{{Fortney}, {Shabram}, {Showman}, {Lian},
  {Freedman}, {Marley}  \& {Lewis}}{{Fortney}
  et~al.}{2010}]{fortney10transmission}
{Fortney} J.~J.,  {Shabram} M.,  {Showman} A.~P.,  {Lian} Y.,  {Freedman}
  R.~S.,  {Marley} M.~S.,   {Lewis} N.~K.,  2010, \mn@doi [ApJ]
  {10.1088/0004-637X/709/2/1396}, \href
  {http://cdsads.u-strasbg.fr/abs/2010ApJ...709.1396F} {709, 1396}

\bibitem[\protect\citeauthoryear{{Fossati} et~al.,}{{Fossati}
  et~al.}{2010}]{fossati10wasp12metals}
{Fossati} L.,  et~al., 2010, \mn@doi [ApJ] {10.1088/2041-8205/714/2/L222},
  \href {http://cdsads.u-strasbg.fr/abs/2010ApJ...714L.222F} {714, L222}

\bibitem[\protect\citeauthoryear{{Fossati}, {Ayres}, {Haswell}, {Bohlender},
  {Kochukhov}  \& {Fl{\"o}er}}{{Fossati} et~al.}{2013}]{fossati13wasp12}
{Fossati} L.,  {Ayres} T.~R.,  {Haswell} C.~A.,  {Bohlender} D.,  {Kochukhov}
  O.,   {Fl{\"o}er} L.,  2013, \mn@doi [ApJ] {10.1088/2041-8205/766/2/L20},
  \href {http://cdsads.u-strasbg.fr/abs/2013ApJ...766L..20F} {766, L20}

\bibitem[\protect\citeauthoryear{{Fuhrmeister}, {Lalitha}, {Poppenhaeger},
  {Rudolf}, {Liefke}, {Reiners}, {Schmitt}  \& {Ness}}{{Fuhrmeister}
  et~al.}{2011}]{fuhrmeister11proxcen}
{Fuhrmeister} B.,  {Lalitha} S.,  {Poppenhaeger} K.,  {Rudolf} N.,  {Liefke}
  C.,  {Reiners} A.,  {Schmitt} J.~H.~M.~M.,   {Ness} J.-U.,  2011, \mn@doi
  [A\&A] {10.1051/0004-6361/201117447}, \href
  {http://cdsads.u-strasbg.fr/abs/2011A%26A...534A.133F} {534, A133}

\bibitem[\protect\citeauthoryear{{Hall}}{{Hall}}{2008}]{hall08chrom}
{Hall} J.~C.,  2008, \mn@doi [Living Reviews in Solar Physics]
  {10.12942/lrsp-2008-2}, \href
  {http://cdsads.u-strasbg.fr/abs/2008LRSP....5....2H} {5}

\bibitem[\protect\citeauthoryear{{Haswell}}{{Haswell}}{2010}]{haswell2010book}
{Haswell} C.~A.,  2010, {Transiting Exoplanets}.
Cambridge University Press, 2010.~ISBN: 9780521139380

\bibitem[\protect\citeauthoryear{{Haswell} et~al.,}{{Haswell}
  et~al.}{2012}]{haswell12}
{Haswell} C.~A.,  et~al., 2012, \mn@doi [ApJ] {10.1088/0004-637X/760/1/79},
  \href {http://cdsads.u-strasbg.fr/abs/2012ApJ...760...79H} {760, 79}

\bibitem[\protect\citeauthoryear{{Henry}, {Marcy}, {Butler}  \& {Vogt}}{{Henry}
  et~al.}{2000}]{henry00hd209458}
{Henry} G.~W.,  {Marcy} G.~W.,  {Butler} R.~P.,   {Vogt} S.~S.,  2000, \mn@doi
  [ApJ] {10.1086/312458}, \href
  {http://cdsads.u-strasbg.fr/abs/2000ApJ...529L..41H} {529, L41}

\bibitem[\protect\citeauthoryear{{Jenkins} et~al.,}{{Jenkins}
  et~al.}{2006}]{jenkins06activity}
{Jenkins} J.~S.,  et~al., 2006, \mn@doi [MNRAS]
  {10.1111/j.1365-2966.2006.10811.x}, \href
  {http://cdsads.u-strasbg.fr/abs/2006MNRAS.372..163J} {372, 163}

\bibitem[\protect\citeauthoryear{{Kramida}, {Reader}, {Ralchenko}  \& {NIST ASD
  Team}}{{Kramida} et~al.}{2015}]{kramida15nist}
{Kramida} A.,  {Reader} J.,  {Ralchenko} Y.,   {NIST ASD Team} 2015, Online:
  http://physics.nist.gov/asd, Online: http://physics.nist.gov/asd

\bibitem[\protect\citeauthoryear{{Lagrange}, {Desort}  \& {Meunier}}{{Lagrange}
  et~al.}{2010}]{lagrange10spots}
{Lagrange} A.,  {Desort} M.,   {Meunier} N.,  2010, \mn@doi [A\&A]
  {10.1051/0004-6361/200913071}, \href
  {http://cdsads.u-strasbg.fr/abs/2010A%26A...512A..38L} {512, A38+}

\bibitem[\protect\citeauthoryear{{Lecavelier Des Etangs}, {Vidal-Madjar},
  {D{\'e}sert}  \& {Sing}}{{Lecavelier Des Etangs}
  et~al.}{2008}]{etangs08rayleigh}
{Lecavelier Des Etangs} A.,  {Vidal-Madjar} A.,  {D{\'e}sert} J.-M.,   {Sing}
  D.,  2008, \mn@doi [A\&A] {10.1051/0004-6361:200809704}, \href
  {http://cdsads.u-strasbg.fr/abs/2008A%26A...485..865L} {485, 865}

\bibitem[\protect\citeauthoryear{{Livingston}, {Wallace}, {White}  \&
  {Giampapa}}{{Livingston} et~al.}{2007}]{livingston07chrom}
{Livingston} W.,  {Wallace} L.,  {White} O.~R.,   {Giampapa} M.~S.,  2007,
  \mn@doi [ApJ] {10.1086/511127}, \href
  {http://cdsads.u-strasbg.fr/abs/2007ApJ...657.1137L} {657, 1137}

\bibitem[\protect\citeauthoryear{{Lodders}}{{Lodders}}{1999}]{lodders99alkali}
{Lodders} K.,  1999, \mn@doi [ApJ] {10.1086/307387}, \href
  {http://cdsads.u-strasbg.fr/abs/1999ApJ...519..793L} {519, 793}

\bibitem[\protect\citeauthoryear{{Louden} \& {Wheatley}}{{Louden} \&
  {Wheatley}}{2015}]{louden15hd189733}
{Louden} T.,  {Wheatley} P.~J.,  2015, \mn@doi [ApJ]
  {10.1088/2041-8205/814/2/L24}, \href
  {http://cdsads.u-strasbg.fr/abs/2015ApJ...814L..24L} {814, L24}

\bibitem[\protect\citeauthoryear{{Meunier} \& {Delfosse}}{{Meunier} \&
  {Delfosse}}{2009}]{meunier09chrom}
{Meunier} N.,  {Delfosse} X.,  2009, \mn@doi [A\&A]
  {10.1051/0004-6361/200911823}, \href
  {http://cdsads.u-strasbg.fr/abs/2009A%26A...501.1103M} {501, 1103}

\bibitem[\protect\citeauthoryear{Noyes, Hartmann, Baliunas, Duncan  \&
  Vaughan}{Noyes et~al.}{1984}]{noyes84}
Noyes R.~W.,  Hartmann L.,  Baliunas S.~L.,  Duncan D.~K.,   Vaughan A.~H.,
  1984, ApJ, 279, 763

\bibitem[\protect\citeauthoryear{{Pont} et~al.,}{{Pont}
  et~al.}{2007}]{pont07hd189733}
{Pont} F.,  et~al., 2007, \mn@doi [A\&A] {10.1051/0004-6361:20078269}, \href
  {http://cdsads.u-strasbg.fr/abs/2007A%26A...476.1347P} {476, 1347}

\bibitem[\protect\citeauthoryear{{Pont}, {Knutson}, {Gilliland}, {Moutou}  \&
  {Charbonneau}}{{Pont} et~al.}{2008}]{pont08hd189733}
{Pont} F.,  {Knutson} H.,  {Gilliland} R.~L.,  {Moutou} C.,   {Charbonneau} D.,
   2008, \mn@doi [MNRAS] {10.1111/j.1365-2966.2008.12852.x}, \href
  {http://cdsads.u-strasbg.fr/abs/2008MNRAS.385..109P} {385, 109}

\bibitem[\protect\citeauthoryear{{Pont}, {Sing}, {Gibson}, {Aigrain}, {Henry}
  \& {Husnoo}}{{Pont} et~al.}{2013}]{pont13hd189733}
{Pont} F.,  {Sing} D.~K.,  {Gibson} N.~P.,  {Aigrain} S.,  {Henry} G.,
  {Husnoo} N.,  2013, \mn@doi [MNRAS] {10.1093/mnras/stt651}, \href
  {http://adsabs.harvard.edu/abs/2013MNRAS.432.2917P} {432, 2917}

\bibitem[\protect\citeauthoryear{{Rauscher} \& {Marcy}}{{Rauscher} \&
  {Marcy}}{2006}]{rauscher06}
{Rauscher} E.,  {Marcy} G.~W.,  2006, \mn@doi [PASP] {10.1086/503021}, \href
  {http://cdsads.u-strasbg.fr/abs/2006PASP..118..617R} {118, 617}

\bibitem[\protect\citeauthoryear{{Redfield}, {Endl}, {Cochran}  \&
  {Koesterke}}{{Redfield} et~al.}{2008}]{redfield08}
{Redfield} S.,  {Endl} M.,  {Cochran} W.~D.,   {Koesterke} L.,  2008, \mn@doi
  [ApJ] {10.1086/527475}, \href
  {http://cdsads.u-strasbg.fr/abs/2008ApJ...673L..87R} {673, L87}

\bibitem[\protect\citeauthoryear{{Seager}}{{Seager}}{2003}]{seager03hd209458}
{Seager} S.,  2003, in {Deming} D.,  {Seager} S.,  eds,  Astronomical Society
  of the Pacific Conference Series Vol. 294, Scientific Frontiers in Research
  on Extrasolar Planets. pp 457--466 (\mn@eprint {} {astro-ph/0305338})

\bibitem[\protect\citeauthoryear{{Seager} \& {Mall{\'e}n-Ornelas}}{{Seager} \&
  {Mall{\'e}n-Ornelas}}{2003}]{seager03transit}
{Seager} S.,  {Mall{\'e}n-Ornelas} G.,  2003, \mn@doi [ApJ] {10.1086/346105},
  \href {http://cdsads.u-strasbg.fr/abs/2003ApJ...585.1038S} {585, 1038}

\bibitem[\protect\citeauthoryear{{Seager} \& {Sasselov}}{{Seager} \&
  {Sasselov}}{2000}]{seager00transmission}
{Seager} S.,  {Sasselov} D.~D.,  2000, \mn@doi [ApJ] {10.1086/309088}, \href
  {http://cdsads.u-strasbg.fr/abs/2000ApJ...537..916S} {537, 916}

\bibitem[\protect\citeauthoryear{{Shkolnik}, {Walker}  \&
  {Bohlender}}{{Shkolnik} et~al.}{2003}]{shkolnik03hd179949}
{Shkolnik} E.,  {Walker} G.~A.~H.,   {Bohlender} D.~A.,  2003, \mn@doi [ApJ]
  {10.1086/378583}, \href {http://cdsads.u-strasbg.fr/abs/2003ApJ...597.1092S}
  {597, 1092}

\bibitem[\protect\citeauthoryear{{Shkolnik}, {Bohlender}, {Walker}  \& {Collier
  Cameron}}{{Shkolnik} et~al.}{2008}]{shkolnik08onoff}
{Shkolnik} E.,  {Bohlender} D.~A.,  {Walker} G.~A.~H.,   {Collier Cameron} A.,
  2008, \mn@doi [ApJ] {10.1086/527351}, \href
  {http://cdsads.u-strasbg.fr/abs/2008ApJ...676..628S} {676, 628}

\bibitem[\protect\citeauthoryear{{Sing} et~al.,}{{Sing}
  et~al.}{2011}]{sing11hd189733}
{Sing} D.~K.,  et~al., 2011, \mn@doi [MNRAS]
  {10.1111/j.1365-2966.2011.19142.x}, \href
  {http://cdsads.u-strasbg.fr/abs/2011MNRAS.416.1443S} {416, 1443}

\bibitem[\protect\citeauthoryear{{Sing} et~al.,}{{Sing}
  et~al.}{2012}]{sing12xo2b}
{Sing} D.~K.,  et~al., 2012, \mn@doi [MNRAS]
  {10.1111/j.1365-2966.2012.21938.x}, \href
  {http://cdsads.u-strasbg.fr/abs/2012MNRAS.426.1663S} {426, 1663}

\bibitem[\protect\citeauthoryear{{Sing} et~al.,}{{Sing}
  et~al.}{2016}]{sing16nature}
{Sing} D.~K.,  et~al., 2016, \mn@doi [\nat] {10.1038/nature16068}, \href
  {http://cdsads.u-strasbg.fr/abs/2016Natur.529...59S} {529, 59}

\bibitem[\protect\citeauthoryear{{Snellen}, {Albrecht}, {de Mooij}  \& {Le
  Poole}}{{Snellen} et~al.}{2008}]{snellen08hd209458}
{Snellen} I.~A.~G.,  {Albrecht} S.,  {de Mooij} E.~J.~W.,   {Le Poole} R.~S.,
  2008, \mn@doi [A\&A] {10.1051/0004-6361:200809762}, \href
  {http://cdsads.u-strasbg.fr/abs/2008A%26A...487..357S} {487, 357}

\bibitem[\protect\citeauthoryear{{Torres}, {Winn}  \& {Holman}}{{Torres}
  et~al.}{2008}]{torres08}
{Torres} G.,  {Winn} J.~N.,   {Holman} M.~J.,  2008, \mn@doi [ApJ]
  {10.1086/529429}, \href {http://cdsads.u-strasbg.fr/abs/2008ApJ...677.1324T}
  {677, 1324}

\bibitem[\protect\citeauthoryear{{Triaud} et~al.,}{{Triaud}
  et~al.}{2009}]{triaud09hd189733}
{Triaud} A.~H.~M.~J.,  et~al., 2009, \mn@doi [A\&A]
  {10.1051/0004-6361/200911897}, \href
  {http://cdsads.u-strasbg.fr/abs/2009A%26A...506..377T} {506, 377}

\bibitem[\protect\citeauthoryear{Vaughan, Preston  \& Wilson}{Vaughan
  et~al.}{1978}]{vaughan78}
Vaughan A.~M.,  Preston G.~W.,   Wilson O.~C.,  1978, PASP, 90, 267

\bibitem[\protect\citeauthoryear{{Vernazza}, {Avrett}  \& {Loeser}}{{Vernazza}
  et~al.}{1981}]{vernazza81}
{Vernazza} J.~E.,  {Avrett} E.~H.,   {Loeser} R.,  1981, \mn@doi [ApJS]
  {10.1086/190731}, \href {http://cdsads.u-strasbg.fr/abs/1981ApJS...45..635V}
  {45, 635}

\bibitem[\protect\citeauthoryear{{Winn} et~al.,}{{Winn} et~al.}{2006}]{winn06}
{Winn} J.~N.,  et~al., 2006, ApJ, 653, L69

\bibitem[\protect\citeauthoryear{{Winn} et~al.,}{{Winn}
  et~al.}{2007}]{winn07hd189733b}
{Winn} J.~N.,  et~al., 2007, AJ, 133, 1828

\bibitem[\protect\citeauthoryear{{Wright}, {Marcy}, {Butler}  \&
  {Vogt}}{{Wright} et~al.}{2004}]{wright04activity}
{Wright} J.~T.,  {Marcy} G.~W.,  {Butler} R.~P.,   {Vogt} S.~S.,  2004, \mn@doi
  [ApJS] {10.1086/386283}, \href
  {http://cdsads.u-strasbg.fr/abs/2004ApJS..152..261W} {152, 261}

\bibitem[\protect\citeauthoryear{{Wyttenbach}, {Ehrenreich}, {Lovis}, {Udry}
  \& {Pepe}}{{Wyttenbach} et~al.}{2015}]{wyttenbach15}
{Wyttenbach} A.,  {Ehrenreich} D.,  {Lovis} C.,  {Udry} S.,   {Pepe} F.,  2015,
  \mn@doi [A\&A] {10.1051/0004-6361/201525729}, \href
  {http://cdsads.u-strasbg.fr/abs/2015A%26A...577A..62W} {577, A62}

\bibitem[\protect\citeauthoryear{{Yan}, {Fosbury}, {Petr-Gotzens}, {Zhao}  \&
  {Pall{\'e}}}{{Yan} et~al.}{2015}]{yan15clv}
{Yan} F.,  {Fosbury} R.~A.~E.,  {Petr-Gotzens} M.~G.,  {Zhao} G.,   {Pall{\'e}}
  E.,  2015, \mn@doi [A\&A] {10.1051/0004-6361/201425220}, \href
  {http://cdsads.u-strasbg.fr/abs/2015A%26A...574A..94Y} {574, A94}

\bibitem[\protect\citeauthoryear{{Zechmeister}, {Anglada-Escud{\'e}}  \&
  {Reiners}}{{Zechmeister} et~al.}{2014}]{zechmeister14}
{Zechmeister} M.,  {Anglada-Escud{\'e}} G.,   {Reiners} A.,  2014, \mn@doi
  [A\&A] {10.1051/0004-6361/201322746}, \href
  {http://cdsads.u-strasbg.fr/abs/2014A%26A...561A..59Z} {561, A59}

\makeatother
\end{thebibliography}






\bsp	
\label{lastpage}
\end{document}